\documentclass[%
 aip,
 amsmath,amssymb,
 reprint,%
]{revtex4-1}

\usepackage{graphicx}
\usepackage{dcolumn}
\usepackage{bm}

\usepackage[utf8]{inputenc}
\usepackage[T1]{fontenc}
\usepackage{mathptmx}
\usepackage{etoolbox}

\usepackage{color} 
\usepackage{subcaption}

\usepackage{ulem}

\makeatletter
\def\@email#1#2{%
 \endgroup
 \patchcmd{\titleblock@produce}
  {\frontmatter@RRAPformat}
  {\frontmatter@RRAPformat{\produce@RRAP{*#1\href{mailto:#2}{#2}}}\frontmatter@RRAPformat}
  {}{}
}%
\makeatother

\newcommand{\ks}{\textcolor{black}} 

\begin{document}

\preprint{AIP/123-QED}

\title{\ks{Dynamics of jet breakup and the resultant drop size distribution: effect of nozzle size and impingement velocity}}
\author{Pavan Kumar Kirar}
\author{Nikhil Kumar}
\author{Kirti Chandra Sahu$^*$}
 \email{ksahu@che.iith.ac.in}
 \affiliation{$^1$Department of Chemical Engineering, Indian Institute of Technology Hyderabad, Sangareddy, 502 284, Telangana, India}

\date{\today}

\begin{abstract}
We conduct systematic experiments to investigate the dynamics of liquid jet breakup and the resulting droplet size distribution, emphasizing the influence of liquid jet velocity and needle exit diameter. We precisely control jet formation using a pressurized water tank equipped with needles of different sizes. Our study quantifies breakup dynamics through dimensionless parameters such as the liquid Weber number and the needle exit area ratio. Our key findings identify three distinct breakup regimes—dripping, Rayleigh, and wind-induced—each dictated by the interplay of surface tension and aerodynamic forces for various combinations of liquid jet velocity and needle exit diameter. We construct a regime map to delineate different breakup behaviours in the $We-A_r$ space. It is observed that lower jet velocities produce narrow probability density functions for jet breakup length due to stable jets, whereas higher velocities result in broader distributions. Increasing jet velocity extends breakup lengths for moderate flow rates due to enhanced stability in the Rayleigh regime, but higher velocities induce instability, leading to shorter breakup lengths. Additionally, we analyze the effects of the needle exit area ratio and liquid Weber number on droplet size distribution, highlighting the transition from mono-modal to bi-modal distribution under varying conditions.
\end{abstract}

\maketitle

\section{Introduction} \label{sec:intro}

Breakup of liquid jets and droplets is relevant in numerous practical applications, including combustion \cite{broumand2016liquid}, spray coating \cite{hashemi2023effects}, pharmaceuticals industries \cite{mousavi2023comparison, wu2019effects}, and printing technologies \cite{lopez1999break, lopez2004note} to name a few. It also plays a vital role in natural phenomena, such as rainfall  \cite{ade2023droplet, ade2023size, soni2020deformation, kirar2022experimental} In these applications, liquid jet atomization is essential, with the primary and secondary breakup of the fluid jet playing a crucial role in determining the efficiency and effectiveness of the atomization process.

In the initial stage of the breakup of a liquid jet, known as primary atomization, a ligament forms and fragments into primary droplets. This stage is followed by secondary atomization, during which these primary droplets further disintegrate into smaller satellite droplets due to aerodynamic forces overcoming viscous and surface tension forces  \cite{faeth1995structure, guildenbecher2009secondary, lefebvre2017atomization}. Different breakup regimes are characterized primarily based on the Weber number $(We)$, which represents the ratio of inertial to surface tension forces. It is defined as $We = \rho U^{2} D/\sigma$, $\rho$ denotes the density of the liquid, $U$ represents the jet velocity, $D$ is the needle diameter and $\sigma$ denotes the interfacial tension of the liquid. The Weber number helps determine whether kinetic or surface tension energy is dominant in the breakup process \cite{lin1998drop, faeth1995structure, hilbing1996droplet}. Various breakup regimes have been identified based on the Weber number, including the Rayleigh regime, the first wind-induced regime, the second wind-induced regime, and the atomization regime \cite{lin1998drop, faeth1995structure, sterling1975instability, miesse1955correlation}.

Previous studies have explored several aspects of liquid jet breakup. \citet{borthakur2019dynamics} numerically investigated the formation and breakup of a liquid jet emerging from an orifice, assessing its behaviour under both gravitational and non-gravitational conditions. They found that the jet breakup length (length between the needle tip and the location where the jet breakup occurs) increases with increasing Weber and Ohnesorge numbers. \ks{The Ohnesorge number is the ratio of viscous forces to inertial and surface tension forces, given by $Oh = \mu/\sqrt{\rho \sigma D}$. Here, $\mu$ represents the liquid viscosity.} \citet{birouk2009liquid} reviewed the disintegration of liquid jets from convergent nozzles, highlighting the effect of various factors, such as internal flow and nozzle geometry, which had been neglected in earlier research. \citet{wu2019effects} examined the impact of bubbles on the breakup of liquid jets, revealing that bubbles significantly reduce the breakup length, thereby enhancing atomization performance. They also noted that the type of gas influences the process, with lighter gases causing a more substantial decrease in breakup length due to fluid momentum conservation. \citet{grant1966newtonian} highlighted the limitations in understanding jet disintegration, noting that existing theories were restricted to low-speed laminar jets in still air. They emphasized the influence of the ambient medium, turbulence, and velocity profile on the jet dynamics. \citet{srinivasan2024primary} also investigated the influence of velocity profile on jet breakup phenomenon. \citet{mousavi2023comparison} investigated the impact of non-Newtonian rheology, specifically shear-thinning behaviour, on jet breakup and droplet production across various regimes. They found that viscosity and rheology significantly influence breakup length and transition in the jetting regime. \citet{tanase2023experimental} examined the initial stages of liquid injection through a capillary nozzle, focusing on the asymmetry of the water-air interface before jet formation. They characterized flow dynamics during drop formation and pinch-off driven by surface tension effects.

Several researchers have also investigated the effect of orifice geometry and injection pressure on liquid jet behaviour. \citet{kiaoulias2019evaluation} assessed the influence of inlet geometry on pressure drop and jet breakup length in single circular orifices in rocket engine injectors. They found that sharp-edge inlets exhibited more significant pressure drop and longer breakup lengths at small orifices, while chamfered inlets showed longer breakup lengths at larger orifice sizes. \citet{sharma2014breakup, wang2015liquid} found that non-circular orifice geometries, such as rectangular, square, and triangular shapes, induce greater instability and faster breakup of low-pressure liquid jets compared to circular orifices of similar cross-sectional areas, highlighting their potential for enhancing fuel atomization and fluid dynamics economically. \citet{borthakur2017formation} investigated liquid drop and jet dynamics during injection through an orifice, examining the interplay of inertia and viscosity. They identified self-similar behaviours in drop growth and the transition from periodic dripping to elongated jetting. \citet{rajesh2016interfacial} presented findings on spray formation from non-circular orifices, revealing superior spray characteristics for elliptical orifices. \citet{geng2020effect} investigated the impact of injection pressure and orifice diameter on soybean biodiesel spray characteristics. They found that higher injection pressures and smaller orifice diameters enhanced atomization. \citet{rajesh2023drop} explored the spray characteristics of kerosene jets from non-circular (elliptic, triangular, square) and circular orifices. Their observations reveal that non-circular orifices exhibited coarsening of atomization, as indicated by an increase in Sauter mean diameter (SMD) across all tested velocities. This was attributed to filament and core breakup, which generated larger liquid structures like filaments and ligaments, particularly noticeable in non-circular orifices. The study also revealed bi-modal drop size distributions at lower velocities, transitioning to mono-modal distributions at higher velocities, with filament breakup resulting in smaller drop sizes in triangular \ks{orifices} than circular orifices.

Few studies have explored the dynamics and instability characteristics of rectangular and elliptical liquid jets. \citet{tadjfar2019effects} conducted a comprehensive review of the literature on rectangular liquid jets, incorporating Rayleigh's model with different aspect ratios and shape type functions. \citet{morad2020axis} numerically investigated low-speed liquid jets from rectangular orifices transitioning into elliptical shapes, observing that axis-switching occurred solely in elliptical cross-sections. They found that higher aspect ratios correlate with shorter breakup lengths due to an increase in the dominant wave number. \citet{jaberi2019wavelength} studied axis-switching oscillations in rectangular and elliptical liquid jets with different aspect ratios. Their model improved predictive accuracy by considering the effects of shape, aspect ratio, and Weber number, incorporating a logarithmic correction term. \citet{kooij2018determines} demonstrates that drop size in the breakup of liquid sheets is primarily determined by the interplay between fluid inertia and surface tension, allowing for predictions based on the Weber number and nozzle geometry. The resulting drop size distribution, described by a compound gamma distribution, varies with nozzle type, with conical nozzles producing more uniform ligaments and flat fan nozzles yielding a broader size range. 

\ks{Table~\ref{tab:table2} provides an overview of previous studies on liquid jet breakup dynamics, categorizing them into experimental and numerical investigations that cover various nozzle geometries, including circular, rectangular, and elliptical shapes, as well as different nozzle diameters. It also highlights the range of Weber ($We$) and Reynolds ($Re$) numbers considered in these studies. However, despite these extensive efforts, a systematic examination of the effects of nozzle diameter and flow rate on different jet breakup dynamics remains unexplored. Our research aims to fill this gap by investigating jet breakup dynamics using circular orifices with eight distinct diameters. In addition to analyzing the breakup phenomenon, we examine the resulting droplet size distribution by varying the liquid jet velocity and needle exit diameter. The results from numerous experiments are presented using dimensionless numbers such as the Weber number ($We$) and the exit area ratio of the nozzle ($A_r$).}

\begin{table*}
\caption{\ks{Summary of earlier studies including the present investigation on jet breakup dynamics.}}
\label{tab:table2}
\begin{ruledtabular}
\begin{tabular}{p{1cm}p{2cm}p{2.3cm}p{2.9cm}p{2cm}p{2cm}p{2cm}}
Reference	&	Approach	&	Jet's liquid	&	Nozzle shape	&	Nozzle diameter (mm)	&	We	&	Re	\\ \hline
Present study	&	Experimental	&	Water	&	Circular	&	0.051 to 0.838	&	0 to 6113.3	&	3.6 to 24586	\\
Ref. \cite{mousavi2023comparison}	&	Numerical	&	Blood	&	Circular	&	1, 1.6, 2.159	&	2 to 6	&	6.45 to 178.87	\\
Ref. \cite{hilbing1996droplet}	&	Numerical	&	-	&	Circular	&	0.283	&	50, 100	&	-	\\
Ref.  \cite{tanase2023experimental}	&	Experimental and numerical	&	-	&	Circular	&	1.8	&	10	&	1,170	\\
Ref.  \cite{wu2019effects}	&	Experimental	&	Water	&	Circular	&	3	&	-	&	-	\\
Ref.  \cite{geng2020effect}	&	Experimental	&	Soybean biodiesel	&	Circular	&	0.366, 0.315, 0.260	&	-	&	-	\\
Ref.  \cite{roth2021high, rezayat2021high}	&	Experimental	&	Water	&	Circular	&	0.6	&	-	&	1790 to 61000	\\
Ref. \cite{kang2023effect}	&	Experimental	&	Water	&	Circular	&	0.495, 0.790	&	4.53 to 38.23	&	-	\\
Ref. \cite{zhan2020effects}	&	Experimental	&	Water, glycerol, ethanol 	&	Circular	&	1, 2, 4	&	-	&	-	\\
Ref. \cite{farvardin2013numerical}	&	Numerical	&	Water	&	Circular, elliptical	&	0.5	&	15 to 330	&	837 to 3884	\\
Ref. \cite{morad2020axis}	&	Numerical	&	Water	&	Circular, rectangle, square, elliptical 	&	0.3 (circular)	&	4 to 80	&	250 to 1000	\\
Ref. \cite{sharma2014breakup}	&	Experimental	&	Water	&	Circular, rectangle, square, triangle 	&	0.31 (circular)	&	0 to 25000	&	-	\\
Ref. \cite{wang2015liquid}	&	Experimental	&	Water	&	Circular, rectangle, square, triangle 	&	0.31 (circular)	&	0 to 2000	&	0 to 6500	\\
Ref. \cite{rajesh2023drop}	&	Experimental	&	Jet A-1	&	Circular, square, triangle, elliptical	&	1.06 (circular)	&	14198 to 38445	&	-	\\
Ref. \cite{rajesh2016interfacial}	&	Experimental	&	Water, glycerol, jet A-1	&	Square, triangle, elliptical	&	4.82 to 4.97	&	-	&	-	\\
Ref. \cite{tadjfar2019effects}	&	Experimental	&	Water	&	Rectangle	&	1.75 to 2.78	&	0 to 1600	&	0 to 25000	\\
Ref. \cite{jaberi2019wavelength}	&	Experimental	&	Water	&	Rectangle, elliptical	&	1.75 to 2.59	&	3 to 1000	&	300 to 20000	\\
\end{tabular}
\end{ruledtabular}
\end{table*}

As discussed above, despite numerous studies, gaps persist in our understanding of jet breakup dynamics. Additionally, only a limited number of researchers have systematically explored jet breakup phenomena. These factors motivated us to undertake a systematic experimental study involving variations in the flow rate of the liquid jet and the needle diameter. In the present study, we designed and conducted experiments to investigate the dynamics of liquid jet breakup and droplet size distribution by varying jet velocity and needle exit diameter. Our experimental setup included a pressurized water dispensing system and a high-speed camera, which captured the jet behaviour, facilitated by a shadowgraph imaging system for detailed analysis. Water served as the working fluid, and we achieved precise control over jet velocity by adjusting the air pressure supplied to the water tank. Needles of varying sizes were utilized to create different jet diameters, and we systematically recorded the jet breakup process at 5000 frames per second (fps). Our analysis involved calculating dimensionless parameters such as the liquid Weber number and needle exit area ratio to characterize the effects of jet velocity and needle diameter on jet breakup length and stability. A regime map was plotted to demarcate different breakup behaviours in the $We-A_r$ space. 

The remainder of the manuscript is organized as follows: Section \ref{sec:Exp}  provides a detailed description of the experimental methodology, covering setup details, varied parameters, data acquisition techniques, post-processing methods, and jet characterization. Section \ref{sec:Res} discusses the results. Finally, concluding remarks are presented in Section \ref{sec:Conc}.

\section{Experimental setup and procedure} \label{sec:Exp}

\begin{figure*}
\centering
     \begin{subfigure}[b]{0.6\textwidth}
         \centering
         \caption{}
         \includegraphics[width=\textwidth]{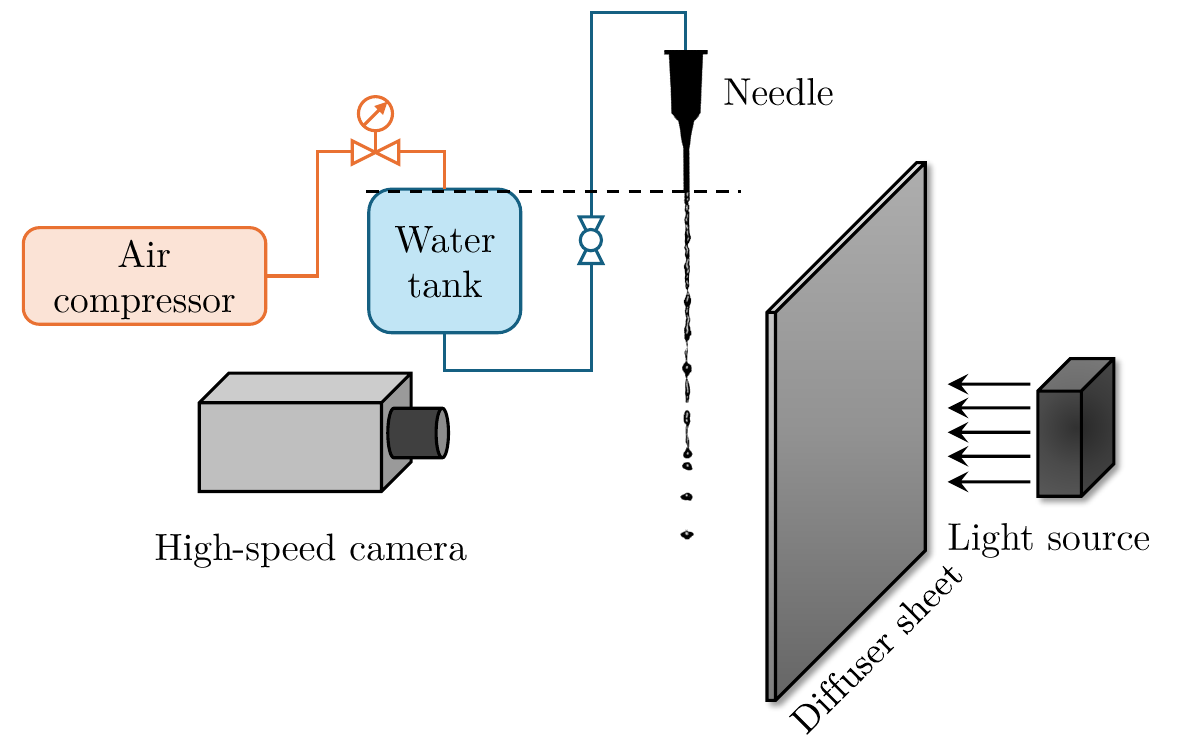}
         \label{fig1a}
     \end{subfigure}
     \hspace{1cm}
     \begin{subfigure}[b]{0.1045\textwidth}
         \centering
         \caption{}
         \includegraphics[width=\textwidth]{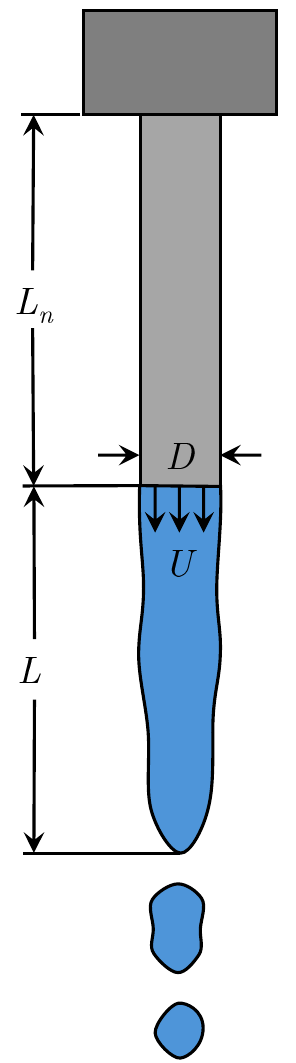}
         \label{fig1b}
     \end{subfigure}
\caption{(a) Schematic representation of the experimental setup. It consists of a water dispensing system and a shadowgraph image-capturing system. (b) A schematic depicting the needle and the jet breakup length. Here, $L_n$ denotes the length of the needle, $D$ represents the diameter of the needle, $U$ is the jet velocity at the needle exit, and $L$ signifies the instantaneous jet breakup length. \ks{The top of the tank and the needle exit are at the same height marked by the horizontal dashed line.}}
\label{fig1:exp}
\end{figure*}

A schematic of the experimental setup used in this study is depicted in \ks{Figure \ref{fig1a}}. It consists of a water dispensing system including an air compressor, a pressurized water tank, a ball valve, a pressure regulator, a dispensing needle, a water collector, and a shadowgraph image-capturing system which uses a high-speed camera, a diffuser sheet, and a light source. Water is used as the working fluid having density, $\rho = 997$ kg/m$^3$, viscosity, $\mu = 0.001$ Pa$\cdot$s, and surface tension, $\sigma = 0.072$ N/m.

The water tank, equipped with an inlet at the top and an outlet at the bottom, is initially filled with deionized (DI) water. An air compressor connected to the inlet supplies pressurized air to the tank, with an air dryer and moisture remover ensuring air quality in the compressed air line. The air pressure within the tank is regulated by a pressure regulator positioned at the inlet line of the water tank. The bottom outlet is connected to a needle via a polyurethane pipe, enabling the dispensing of liquid. A ball valve installed at the outlet controls the liquid flow from the tank. The dispensed water is collected in a container positioned on the ground. To determine the volumetric flow rate of the water, we measure the liquid volume collected in a measuring cylinder and the corresponding time taken. These measurements are essential for calculating the average jet velocity to characterize the liquid jet. We employ a high-speed camera (model: Phantom VEO 640L; make: Vision Research, USA) with Nikkor lens AF Nikkor 135 mm f/2.0D to capture the videos of the liquid jet. The shadowgraph technique used in the present study employs diffused backlit illuminations (GSVITEC light source; model: MultiLED QT; make: GSVITEC, Germany) to illuminate the background for the camera. To ensure appropriate brightness, we utilize high-power (150 W and 12000 lm) light-emitting diode lights diffused by a diffuser sheet as the background light source.

In order to vary the jet diameter, eight needle sizes ranging from 34 gauge to 18 gauge were used. The velocity of the jets was controlled by adjusting the pressurized air supplied to the water tank, with the water tank pressure ($P$) ranging from 10 to 120 psi. Each needle size is characterized by its exit area ratio ($A_r$), defined as $A_r = (D/D_s)^2$, where $D$ is the needle exit diameter and $D_s$ is the exit diameter of the smallest needle \ks{(refer to Table \ref{tab:table1})}. The liquid Weber number, defined as $We = \rho U^{2} D/\sigma$, where $\rho$ is the density of water, $U$ is the jet velocity at the needle exit, $D$ is the needle exit diameter, and $\sigma$ is the interfacial surface tension of water. In the present study, the dynamics of the jet is characterized by varying $A_r$ and $We$. 

\begin{table*}
\caption{The experimental parameters used in the present study. The columns include symbols, the needle gauge, needle diameter ($D$), needle exit area ratio ($A_r$), and liquid Weber number ($We = \rho U^{2} D/\sigma$). The increasing sizes of circular symbols correspond to increasing needle exit diameters, a convention maintained consistently throughout the manuscript.}
\label{tab:table1}
\begin{ruledtabular}
\begin{tabular}{ccccc}
~~Case~~	& ~Needle size (gauge)~	&	~~$D$ (mm)~~	&	~~$A_r$~~	&	~~$We$~~ \\ \hline
        1      & 34	&	\ks{0.051}	&	1	&	0.002 - 0.288  \\
        2	&	32	&	\ks{0.108}	&	4.5	&	0.06 - 4.08    \\
        3	&	 30	&	\ks{0.159}	&	10	&	0.6 - 21.2     \\
        4	&	27	&	0.21	&	17	&	1.4 - 35.0     \\
        5	&	25	&	0.26	&	26	&	6.1 - 141.9    \\
        6	&	22	&	\ks{0.413}	&	66	&	39.2 - 982.3   \\
        7	&	 20	&	\ks{0.603}	    &	140	&	280.3 - 3780.7 \\
        8	&	 18	&	\ks{0.838}	&	~~270~~	&	~~398.3 - 6113.3~~\\
\end{tabular}
\end{ruledtabular}
\end{table*}

The experiment began by filling the water tank with deionized (DI) water until the water level aligned with the needle exit. The needle exit and the top of the water tank were positioned in a straight line. An air pipe from the compressor line was connected to the top of the water tank, and the desired pressure was set using a pressure regulator. The ball valve was fully opened to pressurize the water tank, causing the water jet to emerge from the needle into the quiescent air. After a stabilization period of 5 seconds, a high-speed camera was triggered to record the jet and its subsequent breakup. Due to the rapid dynamics involved, all experiments were recorded at 5000 frames per second (fps) with a resolution of 400 $\times$ 2560 pixels, covering an area of 20.38 mm $\times$ 130.41 mm. Each frame had an exposure time of 1 microsecond ($\mu$s), and the spatial resolution was 19.63 pixels per millimetre (pix/mm). The recorded videos were saved to a local computer. A background video without the water jet was also recorded to assist with post-processing. The Phantom Camera Control (PCC) software was utilized to convert the videos into image sequences. These images were subsequently analyzed using Fiji ImageJ and \textsc{Matlab}$^{\circledR}$ software for further investigation and data extraction.

Figure \ref{fig1b} depicts a schematic of the needle and the emerging liquid jet from the nozzle. The parameter $L_n$ represents the length of the needle, which is maintained constant at 13 mm for all needles used in this study. The parameter $D$ denotes the diameter of the needle, a variable parameter manipulated in this study and reflected in the needle exit area ratio ($A_r$). The parameter $L$ signifies the instantaneous breakup length of the liquid jet ejected from the needle, which varies over time. The jet breakup length was measured from the needle exit tip to the point where the jet breaks up into ligaments and droplets. 

\begin{figure}
\centering
\includegraphics[width=0.45\textwidth]{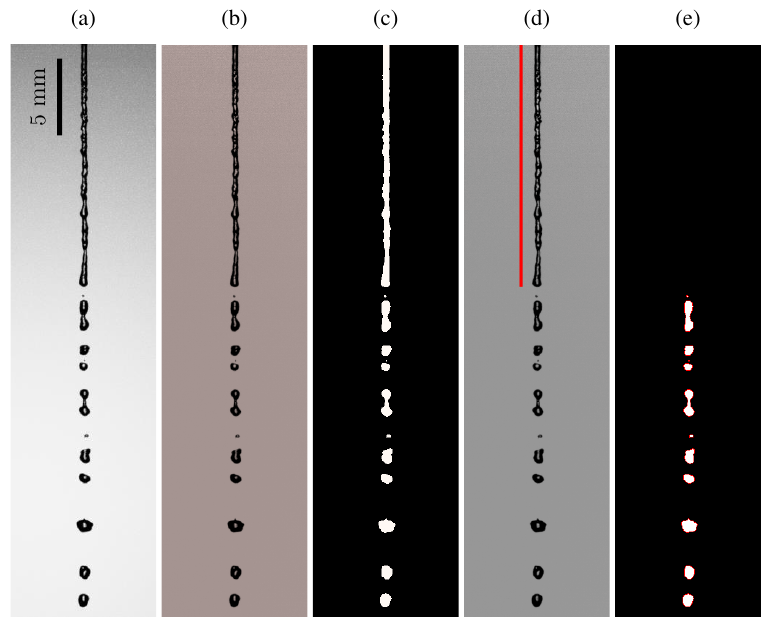}
\caption{Various post-processing steps in the MATLAB\textsuperscript{\textregistered} framework to obtain the jet length and droplet size distribution. The typical case considered here is for $A_r = 66$ and $We = 39.2$. Here, (a) Raw image, (b) Image after background normalization, (c) Image after binarisation, (d) Jet length calculation, and (e) Jet removal and droplet size capturing.}
\label{fig2:matlab}
\end{figure}

The various post-processing steps within the MATLAB\textsuperscript{\textregistered} framework are demonstrated in Figure \ref{fig2:matlab}, involving five sequential steps. Initially, videos are captured using a Photron Camera Controller (PCC), from which individual frames are extracted. Figure \ref{fig2:matlab}(a) represents one such raw frame from the recorded video of the case $A_r=66$ and $We = 39.2$. To enhance the image quality, multiple pre-captured Tag Image File Format (TIFF) background images are read and combined to create a single composite background image. This composite image is normalized to fit the standard 8-bit range, ensuring image consistency. The raw images are then processed by dividing them with the background image to eliminate background effects. Invalid pixel values are replaced with the average of valid pixels, and the mean grayscale values of the images are adjusted. The resulting processed images, demonstrating successful background removal, are saved without compression; Figure \ref{fig2:matlab}(b) shows one such processed frame. Subsequently, the background-normalized image is converted into a binary image (Figure \ref{fig2:matlab}(c)). This binarisation step involves thresholding, where pixel values are set to either black or white, simplifying the images to highlight essential features for further analysis. The final two steps include data collection. The binary images are processed further by reading each frame of a multi-TIFF file. For every image, we obtain the region properties data and specify that from the top region (which is the jet), we store the bounding box data and then calculate the breakup length of the jet from the y-coordinates of the bounding box data. A line is drawn along the left edge of the background normalized image (shown in Figure \ref{fig2:matlab}(b)) to verify the jet length for every frame visually, and the processed images are saved as a new TIFF file shown in Figure \ref{fig2:matlab}(d). Finally, in Figure \ref{fig2:matlab}(e), we isolated the jet using the calculated jet length data and then filtered regions in the image based on area, circularity, and solidity criteria to extract the area equivalent diameters for each detected droplet. We validated the droplet detection by overlaying the edges on the original image, achieving highly accurate droplet identification due to the smooth and transparent background. This comprehensive MATLAB\textsuperscript{\textregistered} processing workflow allows for accurately determining jet length and droplet size distribution, ensuring robust data analysis for our study.

\subsection{Jet characterisation} \label{sub:jet}

\begin{figure}
\centering
     \begin{subfigure}[b]{0.4\textwidth}
         \centering
         \caption{}
         \includegraphics[width=\textwidth]{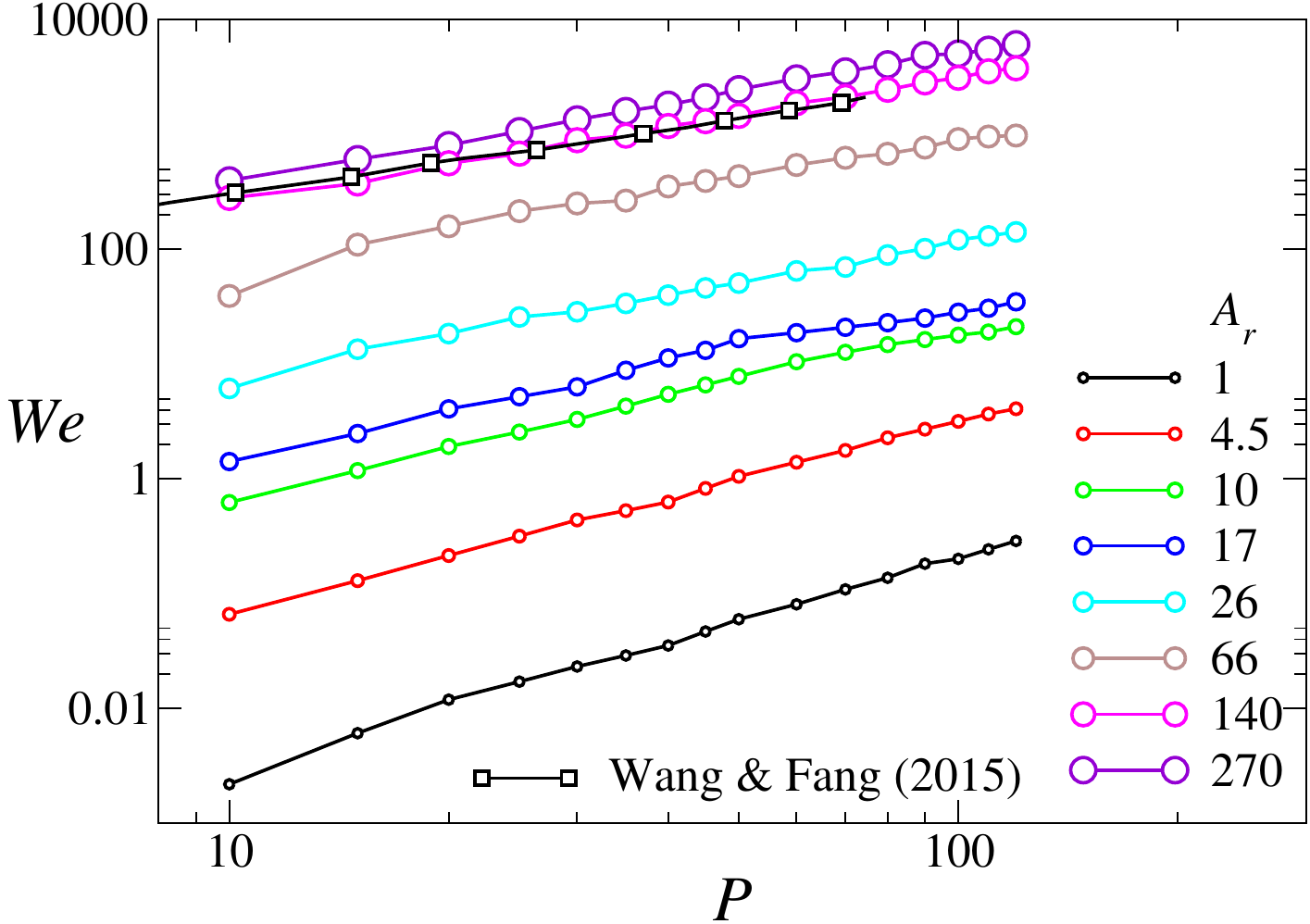}
         \label{fig3a}
     \end{subfigure}
     \hspace{5mm}
     \begin{subfigure}[b]{0.4\textwidth}
         \centering
         \caption{}
         \includegraphics[width=\textwidth]{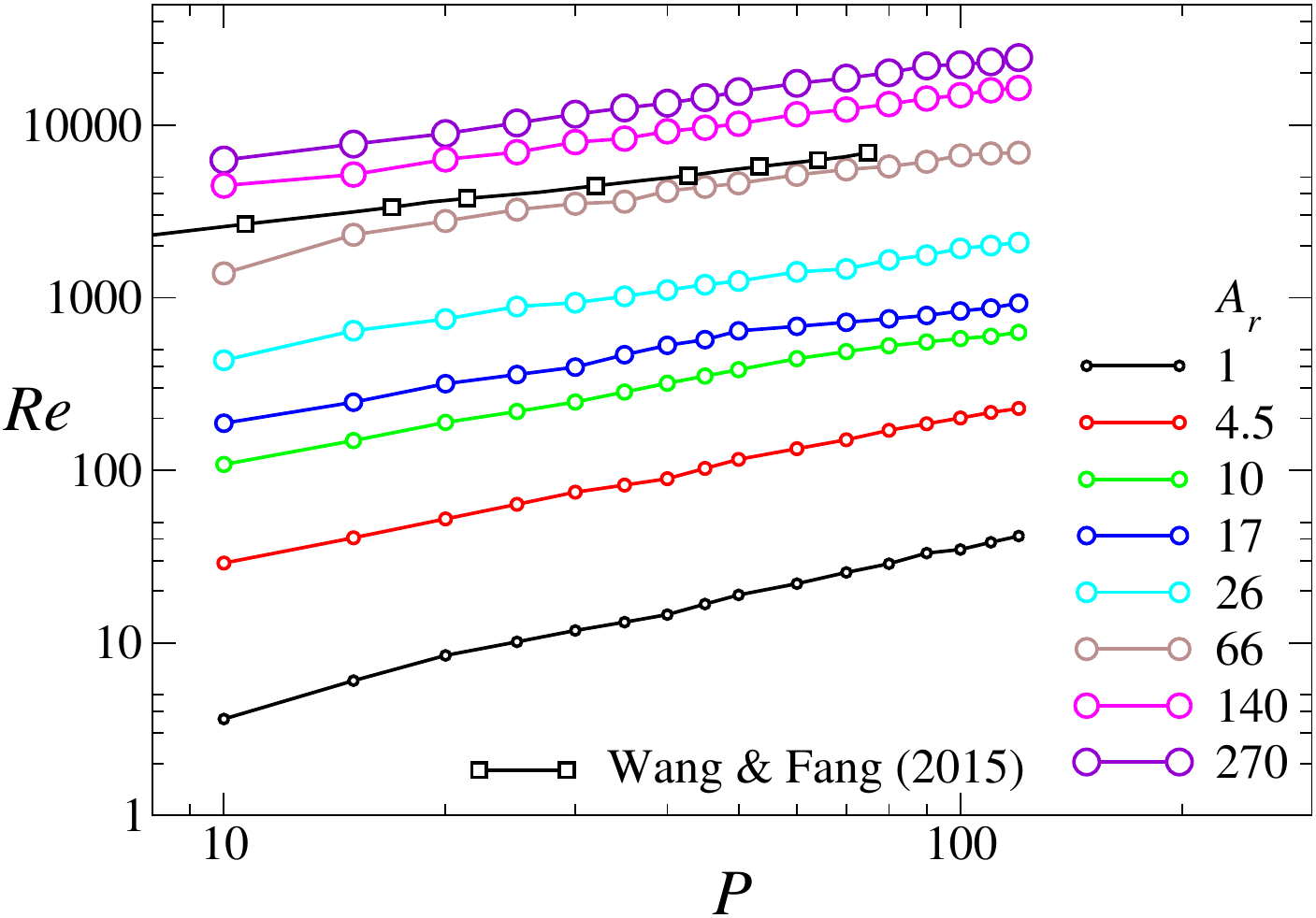}
         \label{fig3b}
     \end{subfigure}
\caption{Variation of (a) liquid Weber number $(We)$, and (b) Reynolds number $(Re)$ with water tank pressure $(P)$ (in psi) for different needle exit area ratios $(A_r)$. The increasing sizes of circular symbols correspond to increasing needle exit diameters. The line with black square symbols shows the results of \citet{wang2015liquid}.}
\label{fig3:WeRe}
\end{figure}

Figure \ref{fig3:WeRe} illustrates the variation in the liquid Weber number ($We = \rho U^{2} D/\sigma$) and Reynolds number ($Re = U \rho D/\mu$) as a function of the water tank pressure ($P$) for various needle exit area ratios ($A_r$) considered in this study. Here, an increase in the size of the circular symbols corresponds to an increase in the needle size (or needle exit area ratio). The line with black squares shows the results of \citet{wang2015liquid} associated with a circular orifice. In our experiments, the water tank pressure ($P$) ranged from 10 to 120 psi, with increments of 5 psi up to 50 psi and 10 psi beyond that. Since deionized (DI) water serves as the constant working fluid in all experiments, its properties, such as density, surface tension, and viscosity, remain unchanged. Therefore, variations in the Weber and Reynolds numbers are solely influenced by the velocity of the liquid jet and the diameter of the needle. For needles with specific exit area ratios, an increase in both the Weber and Reynolds numbers is observed with an increase in water tank pressure, as depicted in Figures \ref{fig3a} and \ref{fig3b}, respectively. This increase correlates with Bernoulli's principle, where higher water tank pressure enhances the liquid mass flow rate, thereby increasing the velocity of the liquid jet. Moreover, at a given water tank pressure, an increase in the Weber and Reynolds numbers is noted as the needle exit area ratio increases, attributable to the simultaneous rise in jet velocity and exit diameter. Since these needles function like capillary tubes with small diameters, increasing the needle exit diameter enhances the liquid mass flow rate and discharge coefficient, resulting in higher jet velocities. Consequently, higher water tank pressures lead to greater volumes of liquid dispensed from the needle over the same duration, further elevating both the Weber and Reynolds numbers.

The characteristics of the jet in our study are compared with findings from \citet{wang2015liquid}, revealing a similar trend where both the Weber and Reynolds numbers increase with pressure, as depicted in Figure \ref{fig3:WeRe}(a) and (b). \citet{wang2015liquid} utilized a circular orifice with a thickness of 2 mm and a diameter of 0.31 mm, resulting in a needle exit area ratio ($A_r$) of 36, which aligns closely with the highest $A_r$ used in our study. This similarity arises despite our longer needle length of 13 mm, as previous research indicates that the needle length relative to its diameter ($L_n/D$) indirectly influences the longevity of the jet dynamics \cite{birouk2009liquid}. The $L_n/D$ ratio in \citet{wang2015liquid} aligns comparably well with our cases ranging from $A_r = 66$ to 270, thus validating the relevance of their findings across these $A_r$ ranges.

\begin{figure}
\centering
\includegraphics[width=0.4\textwidth]{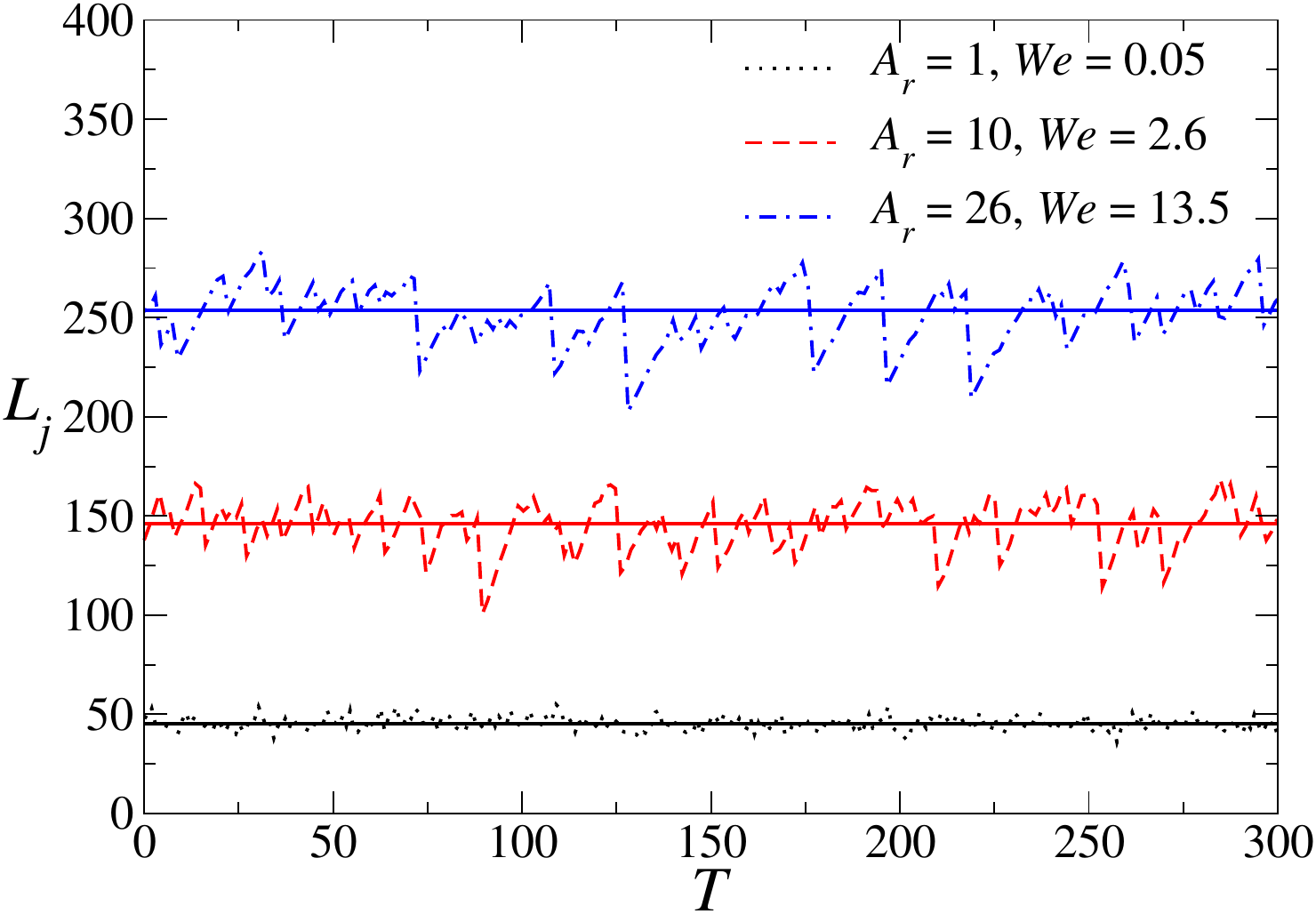}
\caption{Variation of normalised liquid jet breakup length, $L_j=L/D$, with dimensionless  time, $T=tU/D$, for three different cases. Here, $L$ represents the instantaneous jet breakup length, and $t$ denotes physical time in seconds. The black dotted line corresponds to $A_r = 1$, $We=0.05$; the red dashed line is for $A_r = 10$, $We=2.6$; and the blue dot-dash line corresponds to $A_r = 26$, $We=13.5$. Solid straight lines indicate the mean value of jet breakup length for each case. All three cases exhibit characteristics of the Rayleigh breakup regime.}
\label{fig4:jet}
\end{figure}

Figure \ref{fig4:jet} illustrates the temporal evolution of the normalized liquid jet breakup length. The data is presented for three cases characterized by different needle exit area ratios ($A_r$) and liquid Weber numbers ($We$). These cases are represented in the figure by various line styles: the black dotted line (case 1: $A_r = 1$ and $We = 0.05$), the red dashed line (case 2: $A_r = 10$ and $We = 2.6$), and the blue dot-dash line (case 3: $A_r = 26$ and $We = 13.5$). Additionally, the figure includes solid straight lines representing the average jet breakup length for each case. For quantitative analysis, the liquid jet breakup length ($L_j=L/D$) was normalized by the respective needle exit diameter, and the respective needle exit diameter and the jet exit velocity normalized the time ($T=tU/D$). We recorded a 300-frame video using a high-speed camera operating at 5000 frames per second (fps), resulting in a video duration of 60 milliseconds for each experiment. The figure shows that both the needle exit area ratio and the liquid Weber number significantly influence the stability of jet breakup length. As the needle exit area ratio ($A_r$) increases, more pronounced oscillations in the jet breakup length are observed. Similarly, an increase in the liquid Weber number ($We$) also results in more significant variations in the jet breakup length. This indicates a direct correlation between the complexity of the jet breakup process and the needle exit area ratio ($A_r$) and the Weber number ($We$). The solid black, red, and blue lines in the figure represent the mean jet breakup lengths for the first, second, and third cases, with average normalized lengths of 45.3, 146.3, and 253.7, respectively, providing an apparent reference for the average behaviour over time. The oscillatory nature of the jet breakup length in each case can be attributed to the instabilities in the liquid jet, which are more pronounced at higher area ratios and liquid Weber numbers. Figure \ref{fig4:jet} highlights the intricate relationship between needle geometry and jet breakup behaviour. The results demonstrate that larger needle exit area ratios and higher liquid Weber numbers lead to increased jet breakup lengths and more significant variations over time. These findings provide valuable insights into the underlying mechanisms governing liquid jet breakup and are essential for optimizing needle design in various fluid dynamics applications.

\section{Results and discussion} \label{sec:Res}

\subsection{Different breakup phenomena} \label{ss:break}

\begin{figure}
\centering
\includegraphics[width=0.3\textwidth]{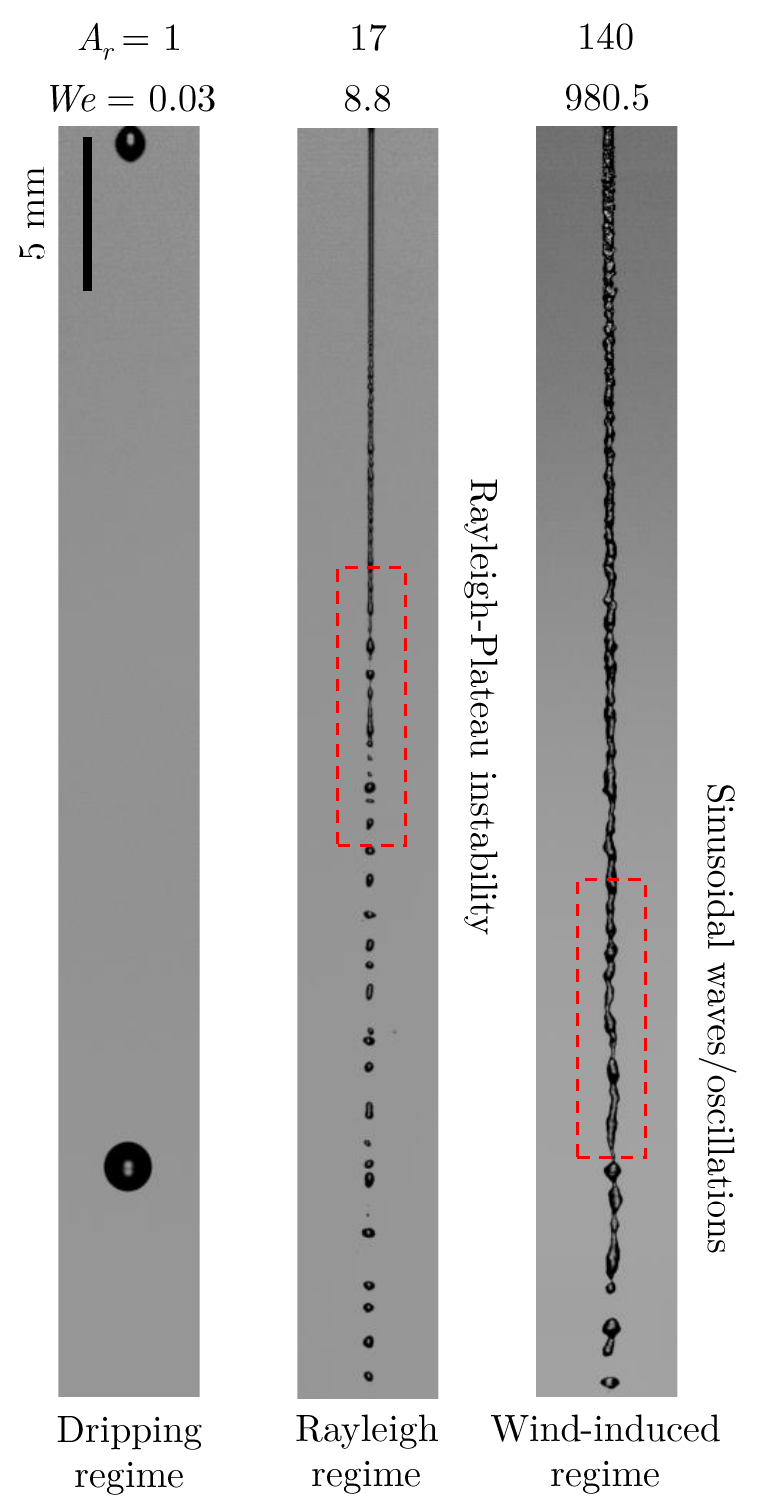}
\caption{Demonstration of various liquid jet breakup regimes, namely dripping, Rayleigh, and wind-induced regimes, observed for different needle exit area ratios and liquid Weber numbers. A scale bar of 5 mm length is included for reference.}
\label{fig5:type}
\end{figure}

The breakup of a liquid jet is influenced by the competition between inertial and aerodynamic forces, leading to distinct breakup regimes depending on jet velocity and liquid properties such as surface tension and viscosity \cite{li2019experimental, chakraborty2022effect, reitz1978atomization, richards1994dynamic, machicoane2023regimes}. As jet velocity increases, breakup transitions from dripping \cite{mousavi2023comparison}, through the Rayleigh regime \cite{roth2021high}, wind-induced regimes \cite{kang2023effect}, and ultimately to atomization \cite{roth2021high, rezayat2021high, dunand2005liquid, daskiran2022impact}. Our study initially explores these observed breakup regimes, followed by an analysis of jet breakup length and droplet size distribution in subsequent subsections. Figure \ref{fig5:type} illustrates these distinct liquid jet breakup regimes, namely dripping, Rayleigh, and wind-induced, corresponding to different needle exit area ratios and liquid Weber numbers, reflecting the breakup behaviours observed under varied flow conditions. In Figure \ref{fig5:type}, a 5 mm scale bar is provided for reference, offering context for the images presented.

In the dripping regime \cite{kovalchuk2018effect, zhan2020effects, furbank2004experimental}, observed at the lowest flow velocities, individual droplets form and dispense one by one from the needle tip without forming a continuous jet. This regime occurs when the liquid Weber number ($We$) is very low, indicating that surface tension dominates over inertial forces. As a result, the liquid does not have sufficient velocity to form a coherent jet, forming discrete droplets due to gravitational and surface tension effects. 

The Rayleigh regime  \cite{rayleigh1879capillary, roth2021high, kalaaji2003breakup} is depicted at slightly higher velocities than the dripping regime but still at relatively low Weber numbers \cite{birouk2009liquid, wang2015liquid, sharma2014breakup, wu2019effects}. In this regime, the jet velocities are low enough that aerodynamic effects are negligible. The breakup of the liquid jet is primarily driven by surface tension forces, which induce axisymmetric disturbances along the jet surface. These disturbances grow in amplitude due to the Rayleigh-Plateau instability until they reach a wavelength approximately equal to the jet radius; at this point, the jet disintegrates into droplets. This regime is characterized by small, periodic surface disturbances and the formation of generally uniform droplets corresponding closely to the diameter of the original jet.

The wind-induced regime \cite{farvardin2013numerical}, also referred to as the sinuous regime \cite{borthakur2017formation, meister1969prediction}, is characterized by longer wavelength waves in the jet \cite{reitz1978atomization}. This regime is observed at intermediate liquid Weber numbers ($We$) \cite{birouk2009liquid, wang2015liquid, sharma2014breakup, wu2019effects}. As the Weber number increases, the wavelengths of these waves progressively shorten \cite{reitz1978atomization}. In this regime, aerodynamic forces play a significant role in the breakup process. The jet forms a wavy appearance due to sinusoidal oscillations around its axis. These oscillations are amplified by aerodynamic forces acting on the jet, leading to its disintegration into droplets. The droplets formed in this regime are comparable in size to the diameter of the jet, and tiny droplets also formed due to the breakup of the threads. The wind-induced regime is distinguished by its higher liquid Weber numbers, indicating that inertial forces are now more significant relative to surface tension forces, causing the jet to break up in a manner driven by both aerodynamic and surface tension effects. Further increases in jet velocity lead to the wind-induced regime with short wavelength waves and, ultimately, atomization. Our experimental setup allows us to achieve a water tank pressure ($P$) of up to 120 psi, resulting in a maximum liquid Weber number of 6113 for an 18-gauge needle. Consequently, we can only achieve the breakup regime up to the wind-induced breakup with long wavelength waves.

\begin{figure}
\centering
\includegraphics[width=0.4\textwidth]{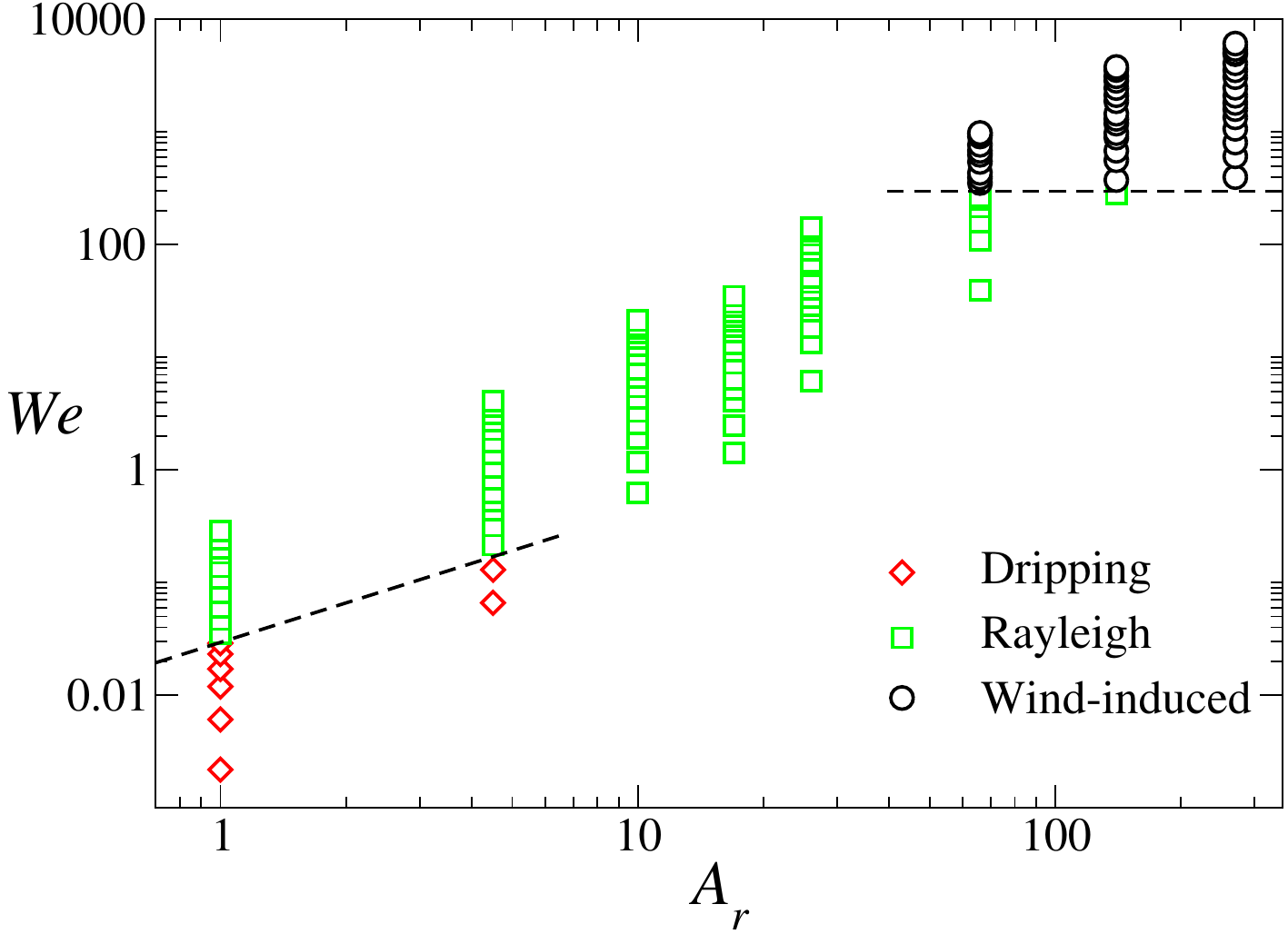}
\caption{Regime map demarcating different breakup behaviours in $We-A_r$ space.}
\label{fig6:regime}
\end{figure}

Figure \ref{fig6:regime} illustrates the complex interaction between two key parameters influencing breakup dynamics: the needle exit area ratio ($A_r$) and the liquid Weber number ($We$). In the figure, red diamonds denote instances of the dripping regime, green squares indicate occurrences of the Rayleigh regime, and black circles represent events within the wind-induced regime. This regime map is essential for understanding how liquid jet breakup modes transition with changes in needle diameter and flow velocity. Note that variations in needle diameter affect $A_r$, thereby altering the Weber number. The figure delineates three distinct flow regimes based on these parameters: dripping occurs at low $A_r$ and low $We$, characterized by steady droplet formation driven by capillary forces. At intermediate $A_r$ and $We$, the Rayleigh regime exhibits periodic instabilities leading to droplet formation, balancing inertial and capillary forces. At high $A_r$ and high $We$, the wind-induced regime dominates, featuring atomization of the fluid stream due to strong aerodynamic interactions and high flow velocities.

\subsection{Jet breakup length}\label{ss:jet}

\begin{figure*}
\centering
     \begin{subfigure}[b]{1.0\textwidth}
         \centering
         \caption{}
         \includegraphics[width=0.9\textwidth]{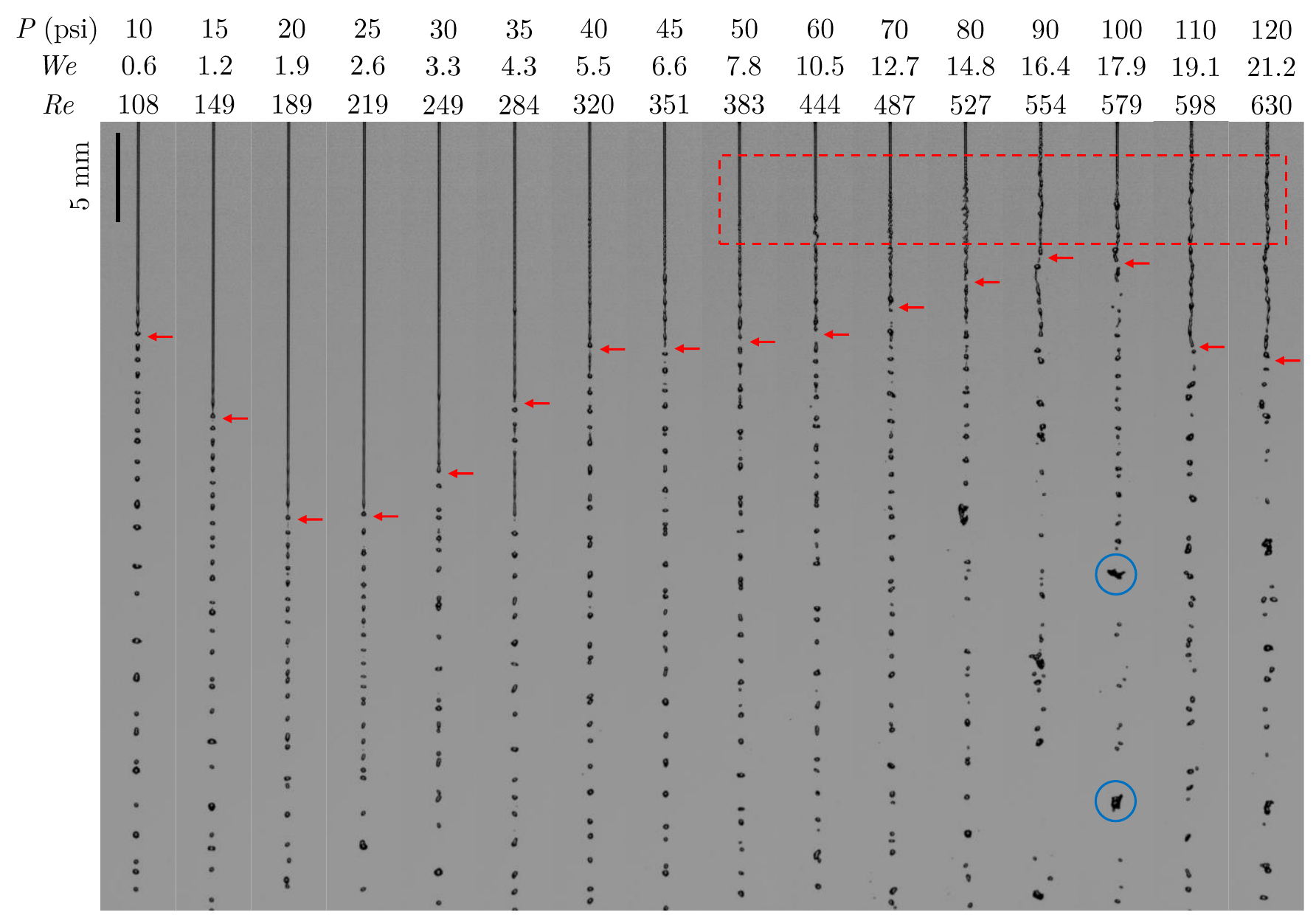}
         \label{fig7a}
     \end{subfigure}
     \vspace{3mm}
     \begin{subfigure}[b]{1.0\textwidth}
         \centering
         \caption{}
         \includegraphics[width=0.9\textwidth]{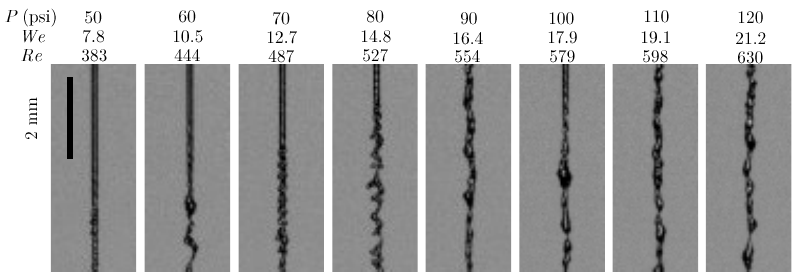}
         \label{fig7b}
     \end{subfigure}
\caption{(a) The breakup of a circular liquid jet across a range of water tank pressures from 10 to 120 psi, with corresponding Weber number values indicated at the top of each panel. The needle exit area ratio is fixed at $A_r =10$. Red arrows denote the location of the jet breakup, while blue circles emphasize the presence of large droplets and ligaments. (b) The enlarged view of the red dashed region in panel (a) illustrates the surface disturbances of the jet at pressures ranging from 50 to 120 psi.}
\label{fig7:pressure}
\end{figure*}

Figure \ref{fig7a} presents the liquid jet breakup length of a circular jet under varying water tank pressures, highlighting the evolution of jet breakup length as pressure increases from 10 to 120 psi, increasing the liquid Weber number from 0.6 to 21.2, \ks{and the Reynolds number from 108 to 630}. The jet breakup length is the distance from the orifice or needle exit to the point where the jet disintegrates into droplets. The red arrows mark the locations where the liquid jet separates, forming discrete droplets or ligaments. Initially, the jet breakup length shows a linear increase with rising pressure, indicative of a laminar regime where the jet remains relatively stable. This linear trend continues up to 25 psi ($We=2.6$ \ks{and $Re=219$}); at this point, the jet is firmly within the Rayleigh breakup regime. Here, the stability of the jet decreases, reaching a peak breakup length before transitioning into a regime characterized by increased instability. As the pressure increases beyond 25 psi ($We=2.6$ \ks{and $Re=219$}), the breakup length decreases, reaching its lowest point around 100 psi ($We=17.9$ \ks{and $Re=579$}). This decrease is due to the jet entering a more unstable phase due to increasing velocity and surface disturbances. The enlarged view of the surface disturbances is shown in Figure \ref{fig7b} from 50 to 120 psi ($We=7.8$ to 21.2 \ks{and $Re=383$ to 630}). However, as the pressure increases to 120 psi ($We=21.2$ \ks{and $Re=630$}), the breakup length increases again, albeit with a smaller slope than the initial linear rise. This non-monotonic behaviour in the jet breakup length is a typical pattern observed in numerous studies \cite{sharma2014breakup, wang2015liquid, birouk2009liquid, grant1966newtonian, zhan2020effects, sun2024effects, kiaoulias2019evaluation, rezayat2021high}. \ks{The phenomenon of the breakup length can be explained as follows: As the Weber number increases, the jet breakup length initially extends in the laminar and Rayleigh regimes due to the dominance of surface tension forces. As the Weber number rises further, the breakup length decreases as the liquid jet transitions to a semi-turbulent state, where turbulent disturbances enhance the breakup process. However, with continued increases in the Weber number, the weakening of surface tension forces allows the jet to stabilize, leading to a subsequent increase in breakup length. A more detailed explanation of this behavior is provided in Figure \ref{fig9:We_Lj}, which illustrates the relationship between the Weber number and jet breakup length.} The surface disturbances and the breakup of the jet occur well within the laminar flow and Rayleigh breakup regime despite showing a different jet length breakup trend. The jet breakup length variation thus illustrates a non-linear relationship between pressure and jet breakup length, with distinct phases corresponding to different breakup regimes. The jet remains stable and laminar at lower pressures, while higher pressures induce instability and more rapid breakup. Beyond a critical pressure, the jet transitions again, showing increasing breakup length. These observations align with findings from other studies \cite{rezayat2021high, birouk2009liquid, grant1966newtonian, zhan2020effects, sun2024effects, kiaoulias2019evaluation}, suggesting that the jet breakup length will continue to increase with further pressure increments beyond 120 psi ($We=21.2$).

\begin{figure*}
\centering
\includegraphics[width=0.9\textwidth]{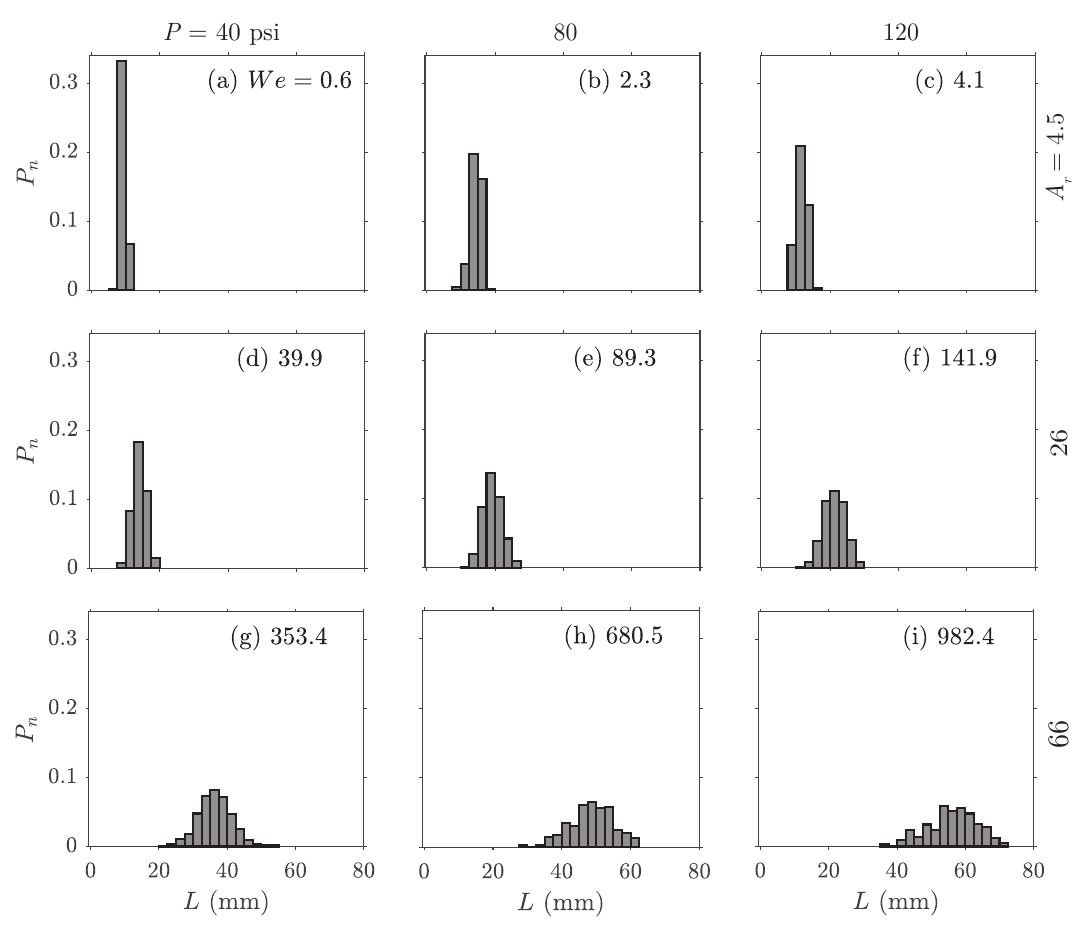}
\caption{Histograms depicting the number probability density function ($P_n$) for jet breakup length, $L$ (mm), over time are shown for nine different cases, illustrating the stability of the jet under varying conditions. Panels (a, b, c), (d, e, f), and (g, h, i) correspond to needle exit area ratios ($A_r$) of 4.5, 26, and 66, respectively. Panels (a, d, g), (b, e, h), and (c, f, i) represent water tank pressures ($P$) of 40, 80, and 120 psi, respectively. Here, the corresponding liquid Weber number ($We$) is also indicated in each panel.}
\label{fig8:pdf1}
\end{figure*}

\begin{figure}
\centering
\includegraphics[width=0.4\textwidth]{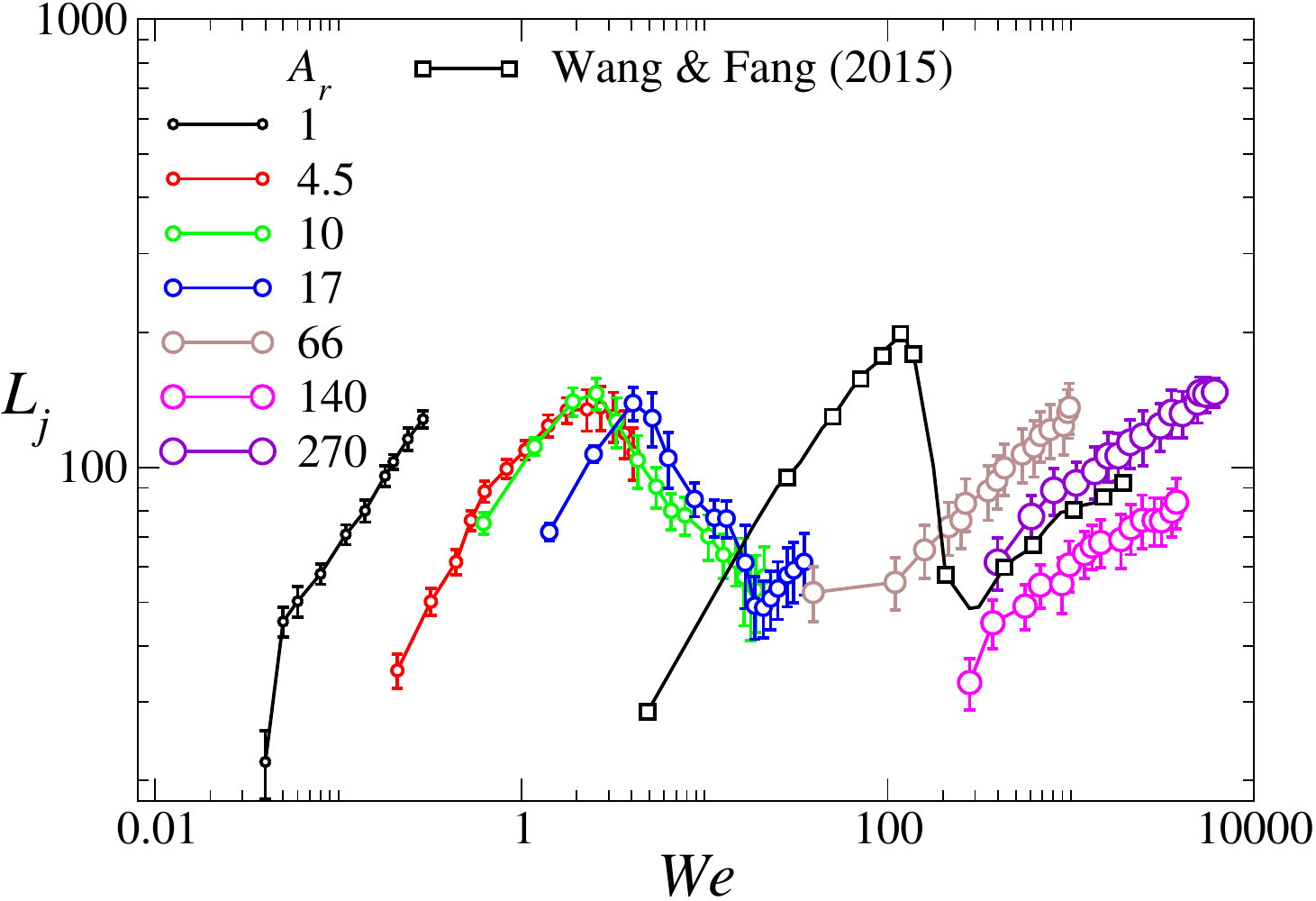}
\caption{Stability curve depicting the dimensionless mean jet breakup length ($L_j = L_{avg}/D$) plotted against the Weber number ($We$) for different needle exit area ratios ($A_r$). Here, $L_{avg}$ represents the mean jet length observed over time. Error bars indicate the temporal variability in $L_j$. The size of the circle symbols increases with larger needle exit diameters, while black squares denote data points sourced from Ref. \cite{wang2015liquid}.}
\label{fig9:We_Lj}
\end{figure}

\begin{figure*}
\centering
\includegraphics[width=0.9\textwidth]{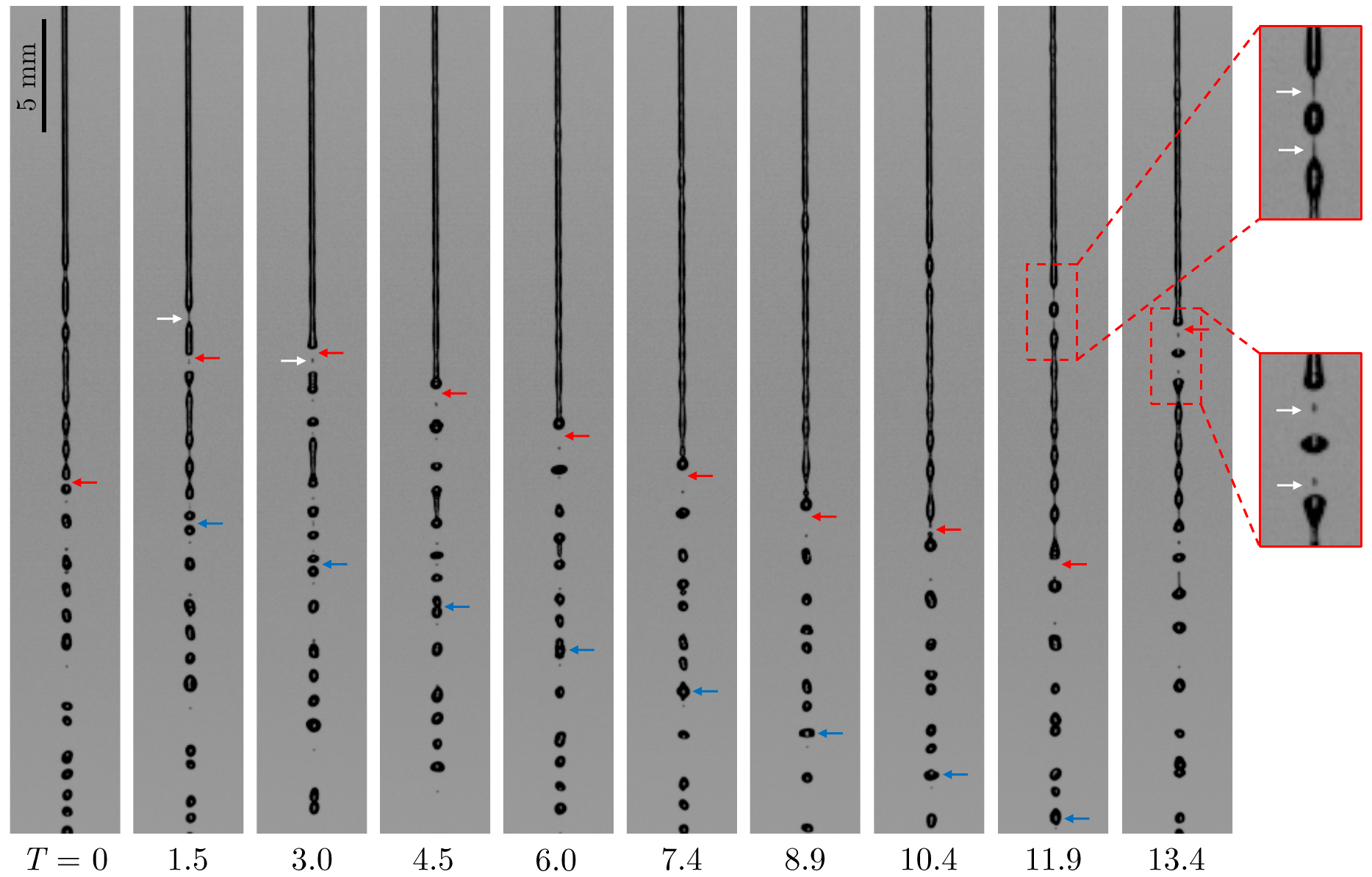}
\caption{Temporal evolution of jet breakup dynamics for $A_r = 26$ and $We = 13.49$. The values of the dimensionless time ($T$) are indicated at the bottom of each image. The red arrows indicate jet breakup points due to Rayleigh-Plateau instability, the blue arrows show droplet coalescence with oscillations, and the white arrows mark the formation of tiny satellite droplets from the liquid thread. Enlarged views depict thread breakup and tiny droplet generation.}
\label{fig10:coalescence}
\end{figure*}

To evaluate the stability of the jet, we analyze the variation in the temporal evolution of the jet breakup length, defined as the distance from the orifice or needle exit to the point where the jet disintegrates into droplets. This parameter is crucial for understanding the jet's stability under varying operating conditions. Figure \ref{fig8:pdf1} displays histograms illustrating the number probability density function ($P_n$) for the instantaneous jet breakup length, $L$ (mm), for different conditions considered in this study. These histograms depict the probability of observing specific ranges of jet breakup lengths over 300 instances, providing insights into the stability of the liquid jet over time. Figure \ref{fig8:pdf1}(a, b, c), Figure \ref{fig8:pdf1}(d, e, f), and Figure \ref{fig8:pdf1}(g, h, i) correspond to needle exit area ratios ($A_r$) of 4.5, 26, and 66, respectively. Panels (a, d, g) represent situations with water tank pressures of $P=40$ psi, panels (b, e, h) of $P=80$ psi, and panels (c, f, i) of $P=120$ psi. Each panel also indicates the corresponding liquid Weber number ($We$). The number probability density function ($P_n$) quantifies the relative likelihood of different breakup lengths occurring within each experimental condition, offering a statistical view of jet breakup stability.

\begin{figure}
\centering
     \begin{subfigure}[b]{0.4\textwidth}
         \centering
         \caption{}
         \includegraphics[width=\textwidth]{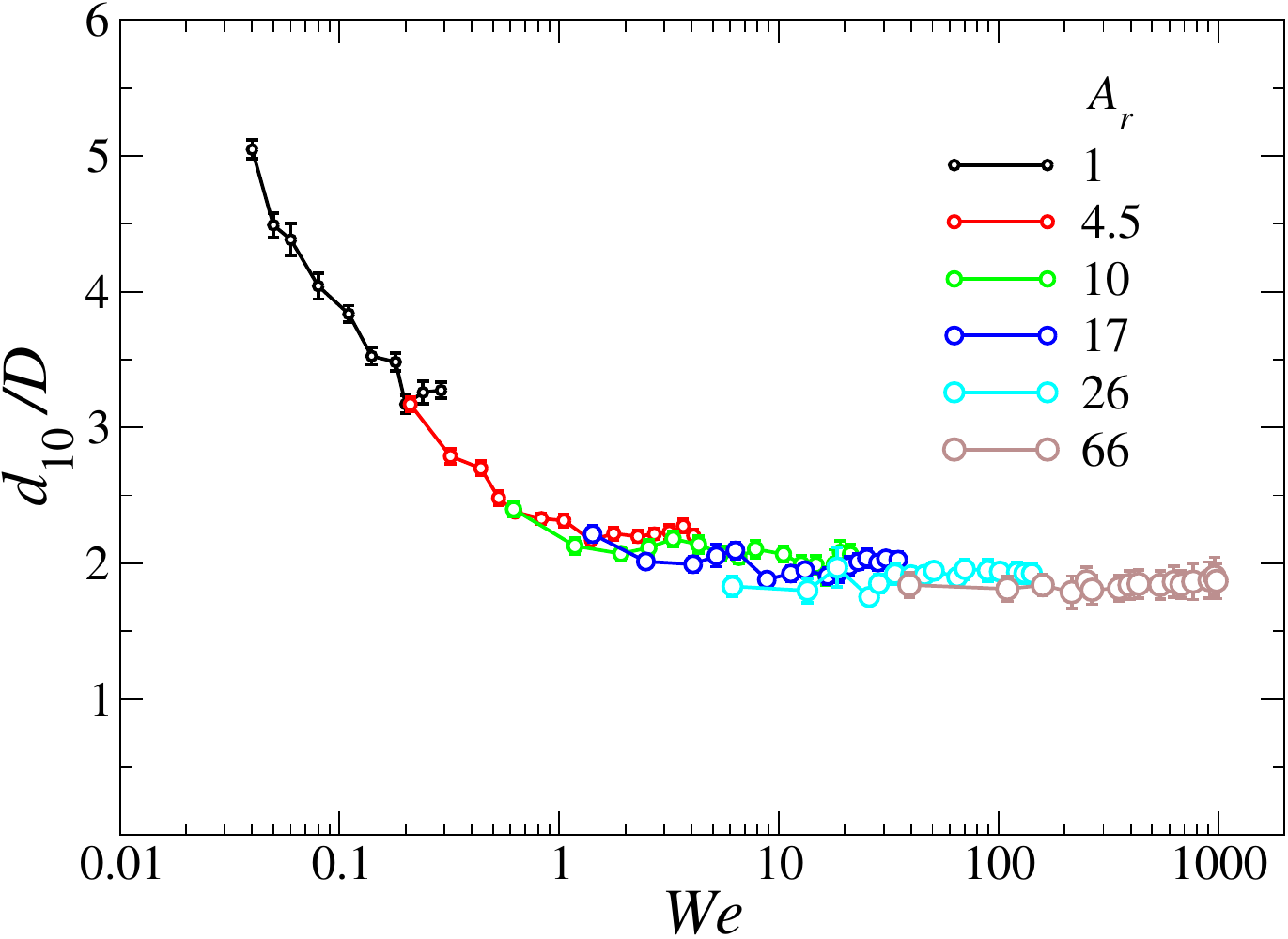}
         \label{fig11a}
     \end{subfigure}
     \hspace{5mm}
     \begin{subfigure}[b]{0.4\textwidth}
         \centering
         \caption{}
         \includegraphics[width=\textwidth]{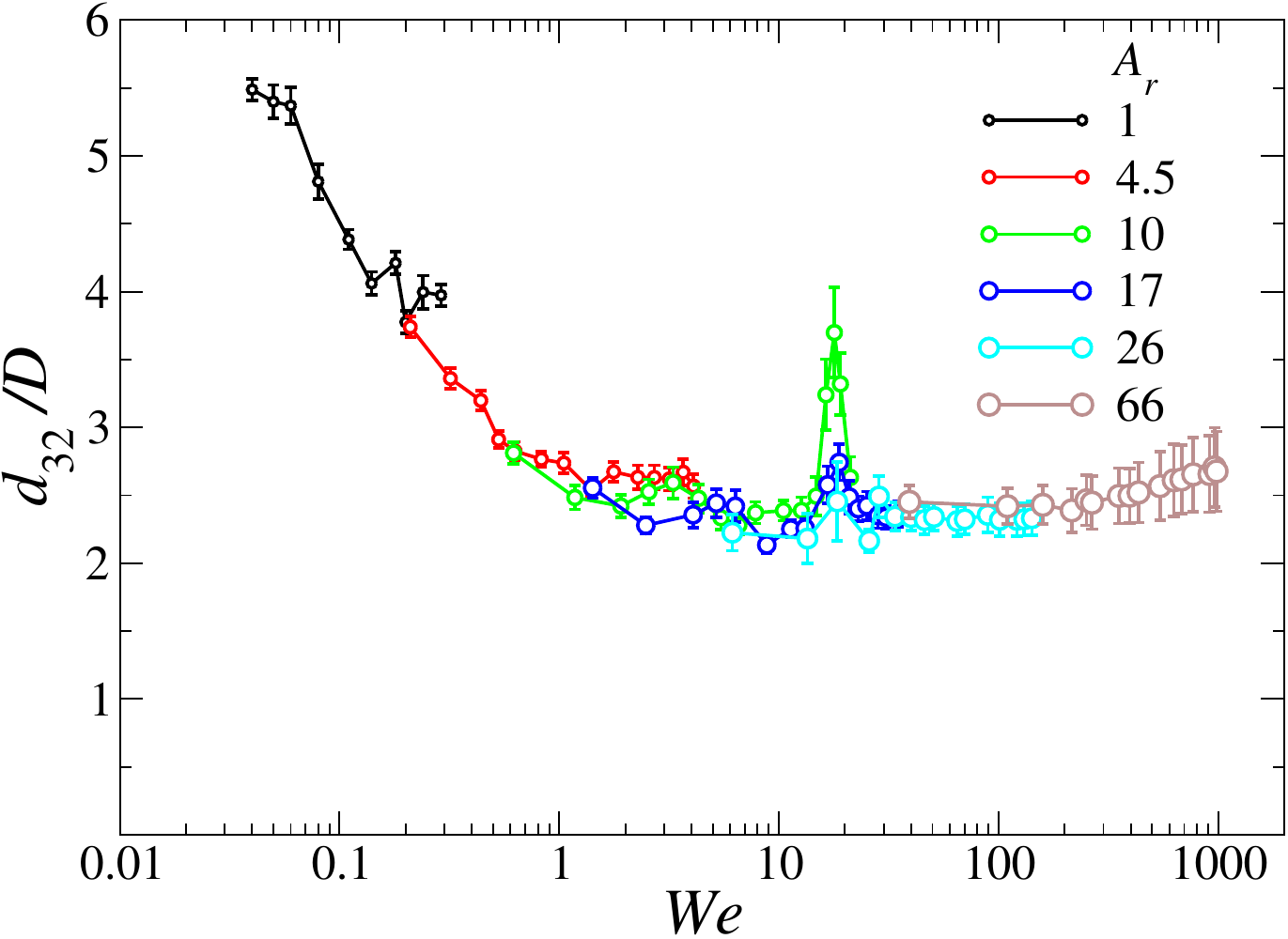}
         \label{fig11b}
     \end{subfigure}
\caption{Variation of (a) normalised number mean diameter ($d_{10}/D$) and (b) Sauter mean diameter ($d_{32}/D$) with the Weber number ($We$) for different values of $A_r$.}
\label{fig11:dropDia}
\end{figure}

Figure \ref{fig8:pdf1}(a) shows the histogram of the probability distribution function (PDF) for the variation in jet breakup length $(L)$ for $A_r=4.5$ and $We=0.6$. The narrow distribution and prominent histogram peak indicate minimal variation in jet breakup length over time, reflected by a lower standard deviation. Here, the average jet breakup length is $L_{avg}=9.54$ mm, with a standard deviation of $L_{std}=0.51$ mm. As the water tank pressure increases, the jet length increases, accompanied by a more significant variation in jet length, as shown in Figure \ref{fig8:pdf1}(b), where $We=2.3$. In this case, the average jet length is $L_{avg}=14.55$ mm with a standard deviation of $L_{std}=1.54$ mm, indicating decreased jet stability with increased pressure. Further increasing the pressure results in a decrease in the average jet length to $L_{avg}=11.66$ mm but with the same standard deviation ($L_{std}=1.55$ mm), as depicted in the histogram for Figure \ref{fig8:pdf1}(c) with $We=4.1$. This reduction in jet length occurs because the jet stability remains within the Rayleigh breakup regime, initiating minimal surface disturbances. Increasing the needle exit area ratio to $A_r = 26$ leads to a substantial increase in the Weber number compared to the previous case ($A_r = 4.5$). As the water tank pressure increases from 40, 80, to 120 psi, the corresponding Weber numbers increase to $We=39.9$, 89.3, and 141.9, respectively, as shown in Figures \ref{fig8:pdf1}(d,e,f). In these cases, the average jet length ($L_{avg}$) increases from 14.05 mm, 19.27 mm, to 21.26 mm, with the standard deviation ($L_{std}$) increasing from 1.96 mm, 2.79 mm, to 3.13 mm. This trend is visually represented by the histograms in Figures \ref{fig8:pdf1}(d, e, f), where the widening of the distribution (increasing width of bars) and a slight decrease in peak height indicate a broader range of jet lengths due to higher mean values and standard deviations. As pressure increases, the jet length extends, following the second slope described in Figure \ref{fig7:pressure}. These Weber number ranges lie between the Rayleigh and wind-induced regimes, signifying a transitional region where jet instability intensifies with higher Weber numbers.

Inspection of the results for $A_r=66$ (Figure \ref{fig8:pdf1}(g-i)) reveals that increasing the pressure leads to an increase in the average jet length ($L_{avg}$) from 36.57 mm to 48.68 mm to 56.25 mm, with the standard deviation ($L_{std}$) rising from 5.13 mm to 6.27 mm to 7.29 mm for $We=353.4$, 680.5, and 982.4, respectively. The decreasing peak height and increasing width of the bars in the histograms further confirm this trend. Observing from top to bottom in Figure \ref{fig8:pdf1}, as the needle exit area ratio increases, the distribution of jet lengths becomes wider due to increased instability in the jet. In Figure \ref{fig8:pdf1}(a-f), it can be seen that the breakup length decreases within the Rayleigh breakup regime, indicating greater stability, as evidenced by the narrower distribution of the histograms. However, with further increases in the Weber number, the regime shifts to the wind-induced regime, as shown by the more spread-out histograms in Figure \ref{fig8:pdf1}(g-i). This observation aligns with findings by \citet{rezayat2021high}, which demonstrate that jet breakup length varies considerably across different regimes. Specifically, the atomization regime exhibits a broader distribution of breakup lengths compared to the Rayleigh and wind-induced regimes, which display less dispersed breakup lengths under lower velocity conditions.

Figure \ref{fig9:We_Lj} illustrates the stability curve for the dimensionless mean jet breakup length ($L_j$) as a function of the Weber number ($We$) for various needle exit area ratios ($A_r$). The error bars denote the temporal variation in jet length. The increasing size of the circle symbols represents increasing needle exit diameters. The black squares indicate the variation in jet length with Weber number as reported by \citet{wang2015liquid}. The stability curve demonstrates a non-monotonic relationship between the jet breakup length and Weber number within the observed regime, consistent with findings from other studies \cite{zhan2020effects, sun2024effects, kiaoulias2019evaluation}. As the water tank pressure, and consequently the liquid Weber number, increases, the jet breakup length grows within the laminar flow and Rayleigh breakup regimes for $A_r=1$ to 17, reaching a maximum length of $L_j = 146.27$. In these conditions, the surface tension force is the primary mechanism driving the breakup, with the jet length peaking in the Weber number range $2.56 \leq We \leq 4.07$. As the Weber number increases further, the jet breakup length begins to decrease. The central portion of the liquid jet transitions to a semi-turbulent state, as previously described and shown in Figure \ref{fig7:pressure}. In this phase, the breakup length diminishes with increasing Weber number until it reaches a minimum value of $L_j=48.65$. This reduction in breakup length continues for cases with $A_r$ from 4.5 to 17, with the decreasing trend observed until the Weber number reaches the range $17.89 \leq We \leq 20.93$ for $A_r=10$ and 17. Turbulent core disturbances enhance the breakup process, reducing the breakup length. Thus, flow turbulence acts as a secondary mechanism influencing the breakup, while surface tension remains the primary factor. It can be observed that for $A_r=10 - 270$, the jet breakup length starts to increase again, but with a smaller slope compared to the initial increase, as the Weber number rises from $We=20.93$. Increasing the Weber number up to 6000 underscores the growing significance of flow turbulence. Numerous researchers have identified liquid turbulence as a critical factor in the wind-induced regime, acting as a primary mechanism for the jet breakup (e.g., Refs. \cite{birouk2009liquid, sharma2014breakup, reitz1978atomization}). As the liquid jet velocity increases, aerodynamic forces also become influential, serving as a secondary mechanism impacting the breakup process. In this regime, the breakup length increases with higher liquid velocity, reaching $L_j=147.2$ in our study. This increase is attributed to the stabilizing effect of turbulence on the breakup process \cite{rezayat2021high, mansour1991dynamic}.

\subsection{Droplet size distribution}

\begin{figure*}
\centering
\includegraphics[width=0.9\textwidth]{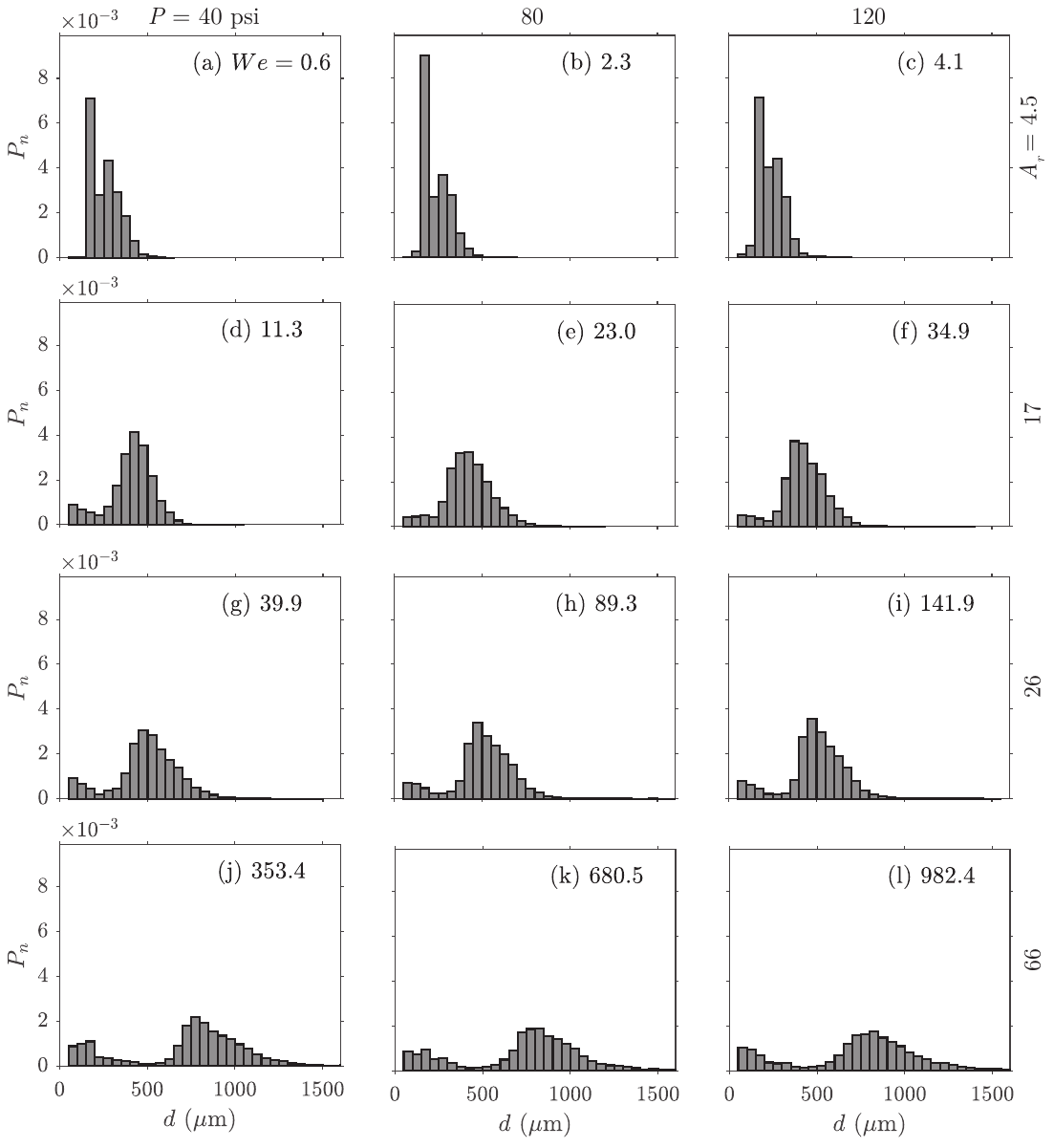}
\caption{Droplet size distribution following the breakup of a liquid jet, depicted as histograms showing the number probability density function ($P_n$) for droplet diameters ($d$) in $\mu$m for twelve different cases. Panels (a, b, c), (d, e, f), (g, h, i), and (j, k, l) correspond to needle exit area ratios ($A_r$) of 4.5, 17, 26, and 66, respectively. Panels (a, d, g, j), (b, e, h, k), and (c, f, i, l) are associated with $P=40$ psi, 80 psi, and 120 psi, respectively. The value of $We$ is indicated in each panel.}
\label{fig12:pdf2}
\end{figure*}

The temporal evolution of a jet breakup phenomenon is presented in Figure \ref{fig10:coalescence} for $A_r = 26$ and $We = 13.49$. This image sequence primarily highlights three critical aspects of the jet breakup process: the jet breakup point (red arrows), the coalescence of droplets (blue arrows), and the formation of tiny droplets from the liquid thread (white arrows). The red arrows in the figure indicate the jet breakup point caused by the Rayleigh-Plateau instability. This phenomenon occurs when surface tension forces lead a cylindrical fluid jet to break into smaller droplets, forming discrete droplets or ligaments. The instability occurs in the Rayleigh regime, where the wavelength of the perturbations on the jet surface dominates the breakup. This Rayleigh breakup process indicates the transition from a continuous jet to separated droplets \cite{javadi2013delayed}. The blue arrows in Figure \ref{fig10:coalescence} highlight the coalescence of droplets. At $T=1.5$, two droplets begin to approach each other and eventually merge by $T=4.5$. The coalesced droplet then oscillates along the impact axis, completing one oscillation cycle from $T=6.0$ to $T=11.9$. During this oscillation, the droplet adopts a prolate shape at $T=6.0$ and $T=11.9$ and an oblate shape at $T=8.9$. This oscillatory behaviour, which reduces the surface area of the merged droplet, is undesirable in many applications. The damping of these oscillations is attributed to viscous dissipation, where internal friction within the fluid gradually diminishes the motion \cite{kirar2022influence, balla2019shape}. In the study by \citet{pal2024collision}, coalescence is prominently displayed in the collision dynamics of drops under specific combinations of velocity and droplet radius ratios, merging two drops into a larger single drop upon collision. According to \citet{viswanathan2019breakup}, simulations reveal that the oscillatory motion of droplets during downstream travel, typically dampened by viscous effects, significantly influences the coalescence and breakup dynamics of primary droplets in the jetting regime. The white arrows in the figure indicate the formation of tiny satellite droplets from the liquid thread. These small droplets form as a result of the breakup of the liquid thread, illustrating the intricate dynamics and fine-scale structures that develop during the jet breakup process. This phenomenon is indicated at $T = 1.5$ and $T = 11.9$, where the Rayleigh-Plateau instability manifests within the jet, resulting in the formation of a thread at the breakup point. The further stretching of the bottom droplets or ligaments separates them from the main jet stream at the top and bottom points of the thread. Consequently, the thread remains and eventually converts into a tiny droplet, as marked at $T = 3.0$ and $T = 13.4$. This breakup process, wherein the jet forms a thread and subsequently a small droplet, significantly influences the droplet size distribution. The detailed effect of this process on droplet size distribution is discussed further in Figure \ref{fig12:pdf2}. Additionally, \citet{kovalchuk2018effect} demonstrated that the transition from dripping to jetting in surfactant-laden liquids occurs at a critical Weber number, underscoring the importance of surface tension in satellite droplet formation. This transition highlights how surface tension plays a crucial role in the dynamics of satellite droplet generation and overall droplet size distribution within the jetting regime.

The number mean diameter ($d_{10}$) and the Sauter mean diameter ($d_{32}$) are crucial metrics for characterizing the distribution of satellite droplets formed during the breakup of a liquid jet. These metrics offer valuable insights into the size distribution of the droplets, which is essential for understanding various jet dynamics and spray processes. The number-based mean diameter ($d_{10}$) represents the arithmetic average of the diameters of all droplets in a sample, weighted by the number of droplets of each diameter. This is given by
\begin{equation}
    d_{10} = \int_{0}^{\infty} d \cdot p(d)~\textrm{d}d,
    \label{eq1}
\end{equation}
where $p(d)$ is the probability density function of the droplet diameters, and $d$ represents the diameter of droplets. This metric provides a straightforward average diameter, giving equal importance to each droplet irrespective of its size.

The Sauter mean diameter (SMD or $d_{32}$) represents the surface area moment mean of the droplet size distribution, considering both the surface area and volume of the droplets. This metric provides a weighted average that emphasizes the surface area, and is given by
\begin{equation}
    d_{32} = \frac{\int_{0}^{\infty} d^3 p(d)~\textrm{d}d}{\int_{0}^{\infty} d^2 p(d)~\textrm{d}d},
    \label{eq2}
\end{equation}

To enable comparison across different needle exit diameters ($D$), both the number mean diameter ($d_{10}$) and the Sauter mean diameter ($d_{32}$) are normalized as $d_{10}/D$ and $d_{32}/D$, respectively, and are plotted against the Weber number ($We$) for various needle exit area ratios ($A_r$) in Figures \ref{fig11a} and \ref{fig11b}. Figure \ref{fig11:dropDia} demonstrates that as the Weber number increases, the normalized number mean and Sauter mean diameters decrease for needle exit area ratios $A_r$ ranging from 1 to 17. Specifically, $d_{10}/D$ decreases from 5 to 2, and $d_{32}/D$ decreases from 5.5 to 2.5 with increasing $We$. This trend indicates that as pressure increases for individual needles, the number-mean and Sauter mean diameters decrease relative to the needle exit diameter $(D)$. The peaks observed for needle exit area ratios $A_r = 10$, 17, and 26 associated with $We = 18$ in Figure \ref{fig11b} coincide with instances where the jet breaks up into droplets or ligaments, forming larger chunks and ligaments. These large chunks or ligaments persist in the flow after the breakup, as indicated by the blue circles in Figure \ref{fig7a}. They travel faster and coalesce with smaller droplets, thereby increasing in size. Consequently, while the volume remains comparable to nearby cases, the surface area of the droplets significantly decreases, resulting in peaks in the Sauter mean diameter. At this intermediate Weber number ($We = 18$), a balance exists between inertial and surface tension forces, leading to a more irregular breakup process where larger droplets and ligaments intermittently form, contributing to the observed peak in the Sauter mean diameter plot. Therefore, the transition between the Rayleigh and transitional regimes preceding the wind-induced regime occurs at $We = 18$ for this set of parameters. In this transitional regime, inertial forces and flow turbulence dominate the breakup process before fully transitioning into the wind-induced regime. For $A_r = 26$ and 66, $d_{10}$ and $d_{32}$ remain relatively constant with increasing Weber number. This constancy occurs because these needle exit area ratios fall within the transitional and wind-induced regimes. Significant changes in droplet diameters occur primarily with larger variations in Weber numbers within these regimes. Thus, for $A_r = 26$ and 66, the Weber number range corresponds to regimes where the number-mean and Sauter mean diameters exhibit minimal variation, unaffected by small changes in Weber number.

Figure \ref{fig12:pdf2} provides the droplet size distribution, illustrating the number probability density function ($P_n$) for droplet diameters after jet breakup for twelve cases. The present study considers the droplet size distribution for twelve distinct cases, as shown in Figure \ref{fig12:pdf2}. The histograms in this figure depict the number probability density function ($P_n$) for droplet diameters ($d$) post-jet breakup. Here, Figure \ref{fig12:pdf2}(a, b, c), (d, e, f), (g, h, i), and (j, k, l), corresponding to $A_r =4.5$, 17, 26, and 66, respectively. In Figure \ref{fig12:pdf2}, panels (a, d, g, j), (b, e, h, k), and (c, f, i, l) are for $P= 40$ psi, 80 psi, and 120 psi, respectively. It can be observed that as the water tank pressure increases (leading to an increase in jet velocity) for all the needle exit area ratios considered, the size distribution shows minimal change because there is no significant variation in the relative Weber number. However, examining changes in needle exit area ratios from top to bottom panels reveals a shift in the peak towards larger droplet diameters, indicating that droplets become larger as the needle diameter increases. Additionally, as the needle exit area ratio increases from top to bottom panels, the Weber number also increases significantly. This increase in the Weber number, indicative of higher jet velocity, leads to the formation of smaller satellite droplets following the thread breakup. These satellite droplets are also discussed in Figure \ref{fig10:coalescence}. Notably, in the case of $A_r=26$, the first peak observed due to tiny droplets starts to form and becomes more pronounced in the case of $A_r=66$. This transition occurs because the needle exit area ratio, $A_r=66$ cases, falls under the wind-induced regime. As the liquid jet velocity rises, aerodynamic forces become significant in this regime, serving as a secondary mechanism that influences the thread breakup process and consequently produces a more substantial number of tiny droplets. In summary, increasing the needle exit area ratio ($A_r$) results in a larger needle exit diameter. At constant water tank pressure, this leads to an increased jet velocity, which in turn raises the Weber number. The increased Weber number causes the breakup regime to transition from Rayleigh to wind-induced, thereby transforming the droplet size distribution from a mono-modal distribution at $A_r=4.5$ and 17 to a bi-modal distribution at $A_r=26$ and 66. This shift highlights the complex interplay between surface tension, which acts as the primary breakup mechanism, and aerodynamic forces, which serve as the secondary breakup mechanism. This underscores the importance of considering these factors when optimizing application atomization processes.

\section{Conclusions} \label{sec:Conc}

We investigate the dynamics of a water jet injected from a needle and the resulting size distribution through systematic experiments that vary jet velocity and needle exit diameter. Our experimental setup includes a pressurized water dispensing system and a high-speed camera to capture jet dynamics, supported by a shadowgraph imaging system for detailed analysis. Three breakup regimes—dripping, Rayleigh, and wind-induced regimes, are identified, each displaying unique dynamics influenced by surface tension and aerodynamic forces. A regime map is constructed to delineate distinct breakup behaviours in the $We-A_r$ space. In the dripping regime, observed at low Weber numbers, individual droplets form and are dispensed one by one from the needle tip without forming a continuous jet. At intermediate Weber numbers, the jet exhibits the Rayleigh breakup, where the Rayleigh-Plateau instability causes the jet to disintegrate into droplets. As the Weber number increases further, the jet displays wind-induced breakup, characterized by significant aerodynamic forces influencing the breakup process. Droplets formed in this regime are comparable in size to the diameter of the jet, with tiny droplets also forming due to the breakup of threads. Our observations indicate a linear increase in jet breakup length with pressure, signifying stable laminar flow up to a Weber number of $We=4.07$. Beyond this threshold, increased pressure induces surface instability, shortening breakup length. At higher pressures, breakup length increases again, marking a transition to the wind-induced regime dominated by aerodynamic effects. The study further reveals that increasing the Weber number decreases the normalized number-mean and Sauter mean diameters. Histograms depicting the probability density function show variations corresponding to jet breakup length, with the droplet size distribution being mono-modal in the Rayleigh regime and bi-modal in the wind-induced regime. It is to be noted that understanding the distribution of droplet sizes following the breakup of a liquid jet is crucial for optimizing atomization processes, which are integral to applications such as fuel injection, agricultural spraying, and pharmaceutical formulations. Analyzing droplet size distribution provides insights into the efficiency of these processes, enabling enhanced performance and effectiveness of spray systems for various industrial applications and natural phenomena.\\

\clearpage
\begin{acknowledgments}
K.C.S. thanks IIT Hyderabad for the financial support through grant IITH/CHE/F011/SOCH1.
\end{acknowledgments}

\section*{AUTHOR DECLARATIONS}
\subsection*{Conflict of Interest}
The authors have no conflicts to disclose.
\subsection*{Author Contributions}
Pavan Kumar Kirar: Investigation (equal); Methodology  (equal); Writing – original draft (equal). Nikhil Kumar: Investigation (equal); Methodology  (equal); Writing – original draft (equal).  Kirti Chandra Sahu: Conceptualization; Supervision; Writing – review \& editing.

\section*{Data Availability Statement}
The data that support the findings of this study are available from the corresponding author upon reasonable request.

\vspace{2mm}


\begin{thebibliography}{57}%
\makeatletter
\providecommand \@ifxundefined [1]{%
 \@ifx{#1\undefined}
}%
\providecommand \@ifnum [1]{%
 \ifnum #1\expandafter \@firstoftwo
 \else \expandafter \@secondoftwo
 \fi
}%
\providecommand \@ifx [1]{%
 \ifx #1\expandafter \@firstoftwo
 \else \expandafter \@secondoftwo
 \fi
}%
\providecommand \natexlab [1]{#1}%
\providecommand \enquote  [1]{``#1''}%
\providecommand \bibnamefont  [1]{#1}%
\providecommand \bibfnamefont [1]{#1}%
\providecommand \citenamefont [1]{#1}%
\providecommand \href@noop [0]{\@secondoftwo}%
\providecommand \href [0]{\begingroup \@sanitize@url \@href}%
\providecommand \@href[1]{\@@startlink{#1}\@@href}%
\providecommand \@@href[1]{\endgroup#1\@@endlink}%
\providecommand \@sanitize@url [0]{\catcode `\\12\catcode `\$12\catcode
  `\&12\catcode `\#12\catcode `\^12\catcode `\_12\catcode `\%12\relax}%
\providecommand \@@startlink[1]{}%
\providecommand \@@endlink[0]{}%
\providecommand \url  [0]{\begingroup\@sanitize@url \@url }%
\providecommand \@url [1]{\endgroup\@href {#1}{\urlprefix }}%
\providecommand \urlprefix  [0]{URL }%
\providecommand \Eprint [0]{\href }%
\providecommand \doibase [0]{http://dx.doi.org/}%
\providecommand \selectlanguage [0]{\@gobble}%
\providecommand \bibinfo  [0]{\@secondoftwo}%
\providecommand \bibfield  [0]{\@secondoftwo}%
\providecommand \translation [1]{[#1]}%
\providecommand \BibitemOpen [0]{}%
\providecommand \bibitemStop [0]{}%
\providecommand \bibitemNoStop [0]{.\EOS\space}%
\providecommand \EOS [0]{\spacefactor3000\relax}%
\providecommand \BibitemShut  [1]{\csname bibitem#1\endcsname}%
\let\auto@bib@innerbib\@empty
\bibitem [{\citenamefont {Broumand}\ and\ \citenamefont
  {Birouk}(2016)}]{broumand2016liquid}%
  \BibitemOpen
  \bibfield  {author} {\bibinfo {author} {\bibfnamefont {M.}~\bibnamefont
  {Broumand}}\ and\ \bibinfo {author} {\bibfnamefont {M.}~\bibnamefont
  {Birouk}},\ }\bibfield  {title} {\enquote {\bibinfo {title} {Liquid jet in a
  subsonic gaseous crossflow: Recent progress and remaining challenges},}\
  }\href@noop {} {\bibfield  {journal} {\bibinfo  {journal} {Prog. Energy
  Combust. Sci.}\ }\textbf {\bibinfo {volume} {57}},\ \bibinfo {pages} {1--29}
  (\bibinfo {year} {2016})}\BibitemShut {NoStop}%
\bibitem [{\citenamefont {Hashemi}\ \emph {et~al.}(2023)\citenamefont
  {Hashemi}, \citenamefont {Shalbaf}, \citenamefont {Jadidi},\ and\
  \citenamefont {Dolatabadi}}]{hashemi2023effects}%
  \BibitemOpen
  \bibfield  {author} {\bibinfo {author} {\bibfnamefont {M.}~\bibnamefont
  {Hashemi}}, \bibinfo {author} {\bibfnamefont {S.}~\bibnamefont {Shalbaf}},
  \bibinfo {author} {\bibfnamefont {M.}~\bibnamefont {Jadidi}}, \ and\ \bibinfo
  {author} {\bibfnamefont {A.}~\bibnamefont {Dolatabadi}},\ }\bibfield  {title}
  {\enquote {\bibinfo {title} {Effects of gas viscosity and liquid-to-gas
  density ratio on liquid jet atomization in crossflow},}\ }\href@noop {}
  {\bibfield  {journal} {\bibinfo  {journal} {AIP Adv.}\ }\textbf {\bibinfo
  {volume} {13}},\ \bibinfo {pages} {035105} (\bibinfo {year} {2023})}\BibitemShut {NoStop}%
\bibitem [{\citenamefont {Mousavi}, \citenamefont {Siavashi},\ and\
  \citenamefont {Bagheri}(2023)}]{mousavi2023comparison}%
  \BibitemOpen
  \bibfield  {author} {\bibinfo {author} {\bibfnamefont {S.}~\bibnamefont
  {Mousavi}}, \bibinfo {author} {\bibfnamefont {M.}~\bibnamefont {Siavashi}}, \
  and\ \bibinfo {author} {\bibfnamefont {M.}~\bibnamefont {Bagheri}},\
  }\bibfield  {title} {\enquote {\bibinfo {title} {Comparison of the jet
  breakup and droplet formation between non-{N}ewtonian and {N}ewtonian
  fluids},}\ }\href@noop {} {\bibfield  {journal} {\bibinfo  {journal} {J.
  Non-Newtonian Fluid Mech.}\ }\textbf {\bibinfo {volume} {321}},\ \bibinfo
  {pages} {105093} (\bibinfo {year} {2023})}\BibitemShut {NoStop}%
\bibitem [{\citenamefont {Wu}\ \emph {et~al.}(2019)\citenamefont {Wu},
  \citenamefont {Zhao}, \citenamefont {Li}, \citenamefont {Xu}, \citenamefont
  {Wang},\ and\ \citenamefont {Liu}}]{wu2019effects}%
  \BibitemOpen
  \bibfield  {author} {\bibinfo {author} {\bibfnamefont {Z.~W.}\ \bibnamefont
  {Wu}}, \bibinfo {author} {\bibfnamefont {H.}~\bibnamefont {Zhao}}, \bibinfo
  {author} {\bibfnamefont {W.~F.}\ \bibnamefont {Li}}, \bibinfo {author}
  {\bibfnamefont {J.~L.}\ \bibnamefont {Xu}}, \bibinfo {author} {\bibfnamefont
  {S.}~\bibnamefont {Wang}}, \ and\ \bibinfo {author} {\bibfnamefont {H.~F.}\
  \bibnamefont {Liu}},\ }\bibfield  {title} {\enquote {\bibinfo {title}
  {Effects of inner bubble on liquid jet breakup},}\ }\href@noop {} {\bibfield
  {journal} {\bibinfo  {journal} {Phys. Fluids}\ }\textbf {\bibinfo {volume}
  {31}},\ \bibinfo {pages} {034107} (\bibinfo {year} {2019})}\BibitemShut {NoStop}%
\bibitem [{\citenamefont {Lopez}, \citenamefont {Soucemarianadin},\ and\
  \citenamefont {Attan{\'e}}(1999)}]{lopez1999break}%
  \BibitemOpen
  \bibfield  {author} {\bibinfo {author} {\bibfnamefont {B.}~\bibnamefont
  {Lopez}}, \bibinfo {author} {\bibfnamefont {A.}~\bibnamefont
  {Soucemarianadin}}, \ and\ \bibinfo {author} {\bibfnamefont {P.}~\bibnamefont
  {Attan{\'e}}},\ }\bibfield  {title} {\enquote {\bibinfo {title} {Break-up of
  continuous liquid jets: effect of nozzle geometry},}\ }\href@noop {}
  {\bibfield  {journal} {\bibinfo  {journal} {J. Imaging Sci. Technol.}\
  }\textbf {\bibinfo {volume} {43}},\ \bibinfo {pages} {145--152} (\bibinfo
  {year} {1999})}\BibitemShut {NoStop}%
\bibitem [{\citenamefont {L{\'o}pez-Herrera}\ and\ \citenamefont
  {Ga{\~n}{\'a}n-Calvo}(2004)}]{lopez2004note}%
  \BibitemOpen
  \bibfield  {author} {\bibinfo {author} {\bibfnamefont {J.~M.}\ \bibnamefont
  {L{\'o}pez-Herrera}}\ and\ \bibinfo {author} {\bibfnamefont {A.~M.}\
  \bibnamefont {Ga{\~n}{\'a}n-Calvo}},\ }\bibfield  {title} {\enquote {\bibinfo
  {title} {A note on charged capillary jet breakup of conducting liquids:
  experimental validation of a viscous one-dimensional model},}\ }\href@noop {}
  {\bibfield  {journal} {\bibinfo  {journal} {J. Fluid Mech.}\ }\textbf
  {\bibinfo {volume} {501}},\ \bibinfo {pages} {303--326} (\bibinfo {year}
  {2004})}\BibitemShut {NoStop}%
\bibitem [{\citenamefont {Ade}\ \emph {et~al.}(2023)\citenamefont {Ade},
  \citenamefont {Kirar}, \citenamefont {Chandrala},\ and\ \citenamefont
  {Sahu}}]{ade2023droplet}%
  \BibitemOpen
  \bibfield  {author} {\bibinfo {author} {\bibfnamefont {S.~S.}\ \bibnamefont
  {Ade}}, \bibinfo {author} {\bibfnamefont {P.~K.}\ \bibnamefont {Kirar}},
  \bibinfo {author} {\bibfnamefont {L.~D.}\ \bibnamefont {Chandrala}}, \ and\
  \bibinfo {author} {\bibfnamefont {K.~C.}\ \bibnamefont {Sahu}},\ }\bibfield
  {title} {\enquote {\bibinfo {title} {Droplet size distribution in a swirl
  airstream using in-line holography technique},}\ }\href@noop {} {\bibfield
  {journal} {\bibinfo  {journal} {J. Fluid Mech.}\ }\textbf {\bibinfo {volume}
  {954}},\ \bibinfo {pages} {A39} (\bibinfo {year} {2023})}\BibitemShut
  {NoStop}%
\bibitem [{\citenamefont {Ade}, \citenamefont {Chandrala},\ and\ \citenamefont
  {Sahu}(2023)}]{ade2023size}%
  \BibitemOpen
  \bibfield  {author} {\bibinfo {author} {\bibfnamefont {S.~S.}\ \bibnamefont
  {Ade}}, \bibinfo {author} {\bibfnamefont {L.~D.}\ \bibnamefont {Chandrala}},
  \ and\ \bibinfo {author} {\bibfnamefont {K.~C.}\ \bibnamefont {Sahu}},\
  }\bibfield  {title} {\enquote {\bibinfo {title} {Size distribution of a drop
  undergoing breakup at moderate weber numbers},}\ }\href@noop {} {\bibfield
  {journal} {\bibinfo  {journal} {J. Fluid Mech.}\ }\textbf {\bibinfo {volume}
  {959}},\ \bibinfo {pages} {A38} (\bibinfo {year} {2023})}\BibitemShut
  {NoStop}%
\bibitem [{\citenamefont {Soni}\ \emph {et~al.}(2020)\citenamefont {Soni},
  \citenamefont {Kirar}, \citenamefont {Kolhe},\ and\ \citenamefont
  {Sahu}}]{soni2020deformation}%
  \BibitemOpen
  \bibfield  {author} {\bibinfo {author} {\bibfnamefont {S.~K.}\ \bibnamefont
  {Soni}}, \bibinfo {author} {\bibfnamefont {P.~K.}\ \bibnamefont {Kirar}},
  \bibinfo {author} {\bibfnamefont {P.}~\bibnamefont {Kolhe}}, \ and\ \bibinfo
  {author} {\bibfnamefont {K.~C.}\ \bibnamefont {Sahu}},\ }\bibfield  {title}
  {\enquote {\bibinfo {title} {Deformation and breakup of droplets in an
  oblique continuous air stream},}\ }\href@noop {} {\bibfield  {journal}
  {\bibinfo  {journal} {Int. J. Multiphase Flow}\ }\textbf {\bibinfo {volume}
  {122}},\ \bibinfo {pages} {103141} (\bibinfo {year} {2020})}\BibitemShut
  {NoStop}%
\bibitem [{\citenamefont {Kirar}\ \emph
  {et~al.}(2022{\natexlab{a}})\citenamefont {Kirar}, \citenamefont {Soni},
  \citenamefont {Kolhe},\ and\ \citenamefont {Sahu}}]{kirar2022experimental}%
  \BibitemOpen
  \bibfield  {author} {\bibinfo {author} {\bibfnamefont {P.~K.}\ \bibnamefont
  {Kirar}}, \bibinfo {author} {\bibfnamefont {S.~K.}\ \bibnamefont {Soni}},
  \bibinfo {author} {\bibfnamefont {P.~S.}\ \bibnamefont {Kolhe}}, \ and\
  \bibinfo {author} {\bibfnamefont {K.~C.}\ \bibnamefont {Sahu}},\ }\bibfield
  {title} {\enquote {\bibinfo {title} {An experimental investigation of droplet
  morphology in swirl flow},}\ }\href@noop {} {\bibfield  {journal} {\bibinfo
  {journal} {J. Fluid Mech.}\ }\textbf {\bibinfo {volume} {938}},\ \bibinfo
  {pages} {A6} (\bibinfo {year} {2022}{\natexlab{a}})}\BibitemShut {NoStop}%
\bibitem [{\citenamefont {Faeth}, \citenamefont {Hsiang},\ and\ \citenamefont
  {Wu}(1995)}]{faeth1995structure}%
  \BibitemOpen
  \bibfield  {author} {\bibinfo {author} {\bibfnamefont {G.~M.}\ \bibnamefont
  {Faeth}}, \bibinfo {author} {\bibfnamefont {L.-P.}\ \bibnamefont {Hsiang}}, \
  and\ \bibinfo {author} {\bibfnamefont {P.-K.}\ \bibnamefont {Wu}},\
  }\bibfield  {title} {\enquote {\bibinfo {title} {Structure and breakup
  properties of sprays},}\ }\href@noop {} {\bibfield  {journal} {\bibinfo
  {journal} {Int. J. Multiphase Flow}\ }\textbf {\bibinfo {volume} {21}},\
  \bibinfo {pages} {99--127} (\bibinfo {year} {1995})}\BibitemShut {NoStop}%
\bibitem [{\citenamefont {Guildenbecher}, \citenamefont {L{\'o}pez-Rivera},\
  and\ \citenamefont {Sojka}(2009)}]{guildenbecher2009secondary}%
  \BibitemOpen
  \bibfield  {author} {\bibinfo {author} {\bibfnamefont {D.~R.}\ \bibnamefont
  {Guildenbecher}}, \bibinfo {author} {\bibfnamefont {C.}~\bibnamefont
  {L{\'o}pez-Rivera}}, \ and\ \bibinfo {author} {\bibfnamefont {P.~E.}\
  \bibnamefont {Sojka}},\ }\bibfield  {title} {\enquote {\bibinfo {title}
  {Secondary atomization},}\ }\href@noop {} {\bibfield  {journal} {\bibinfo
  {journal} {Exp. Fluids}\ }\textbf {\bibinfo {volume} {46}},\ \bibinfo {pages}
  {371--402} (\bibinfo {year} {2009})}\BibitemShut {NoStop}%
\bibitem [{\citenamefont {Lefebvre}\ and\ \citenamefont
  {McDonell}(2017)}]{lefebvre2017atomization}%
  \BibitemOpen
  \bibfield  {author} {\bibinfo {author} {\bibfnamefont {A.~H.}\ \bibnamefont
  {Lefebvre}}\ and\ \bibinfo {author} {\bibfnamefont {V.~G.}\ \bibnamefont
  {McDonell}},\ }\href@noop {} {\emph {\bibinfo {title} {Atomization and
  Sprays}}}\ (\bibinfo  {publisher} {CRC press},\ \bibinfo {year}
  {2017})\BibitemShut {NoStop}%
\bibitem [{\citenamefont {Lin}\ and\ \citenamefont
  {Reitz}(1998)}]{lin1998drop}%
  \BibitemOpen
  \bibfield  {author} {\bibinfo {author} {\bibfnamefont {S.~P.}\ \bibnamefont
  {Lin}}\ and\ \bibinfo {author} {\bibfnamefont {R.~D.}\ \bibnamefont
  {Reitz}},\ }\bibfield  {title} {\enquote {\bibinfo {title} {Drop and spray
  formation from a liquid jet},}\ }\href@noop {} {\bibfield  {journal}
  {\bibinfo  {journal} {Annu. Rev. Fluid Mech.}\ }\textbf {\bibinfo {volume}
  {30}},\ \bibinfo {pages} {85--105} (\bibinfo {year} {1998})}\BibitemShut
  {NoStop}%
\bibitem [{\citenamefont {Hilbing}\ and\ \citenamefont
  {Heister}(1996)}]{hilbing1996droplet}%
  \BibitemOpen
  \bibfield  {author} {\bibinfo {author} {\bibfnamefont {J.~H.}\ \bibnamefont
  {Hilbing}}\ and\ \bibinfo {author} {\bibfnamefont {S.~D.}\ \bibnamefont
  {Heister}},\ }\bibfield  {title} {\enquote {\bibinfo {title} {Droplet size
  control in liquid jet breakup},}\ }\href@noop {} {\bibfield  {journal}
  {\bibinfo  {journal} {Phys. Fluids}\ }\textbf {\bibinfo {volume} {8}},\
  \bibinfo {pages} {1574--1581} (\bibinfo {year} {1996})}\BibitemShut {NoStop}%
\bibitem [{\citenamefont {Sterling}\ and\ \citenamefont
  {Sleicher}(1975)}]{sterling1975instability}%
  \BibitemOpen
  \bibfield  {author} {\bibinfo {author} {\bibfnamefont {A.~M.}\ \bibnamefont
  {Sterling}}\ and\ \bibinfo {author} {\bibfnamefont {C.~A.}\ \bibnamefont
  {Sleicher}},\ }\bibfield  {title} {\enquote {\bibinfo {title} {The
  instability of capillary jets},}\ }\href@noop {} {\bibfield  {journal}
  {\bibinfo  {journal} {J. Fluid Mech.}\ }\textbf {\bibinfo {volume} {68}},\
  \bibinfo {pages} {477--495} (\bibinfo {year} {1975})}\BibitemShut {NoStop}%
\bibitem [{\citenamefont {Miesse}(1955)}]{miesse1955correlation}%
  \BibitemOpen
  \bibfield  {author} {\bibinfo {author} {\bibfnamefont {C.~C.}\ \bibnamefont
  {Miesse}},\ }\bibfield  {title} {\enquote {\bibinfo {title} {Correlation of
  experimental data on the disintegration of liquid jets},}\ }\href@noop {}
  {\bibfield  {journal} {\bibinfo  {journal} {Ind. Eng. Chem. Res.}\ }\textbf
  {\bibinfo {volume} {47}},\ \bibinfo {pages} {1690--1701} (\bibinfo {year}
  {1955})}\BibitemShut {NoStop}%
\bibitem [{\citenamefont {Borthakur}\ \emph {et~al.}(2019)\citenamefont
  {Borthakur}, \citenamefont {Biswas}, \citenamefont {Bandyopadhyay},\ and\
  \citenamefont {Sahu}}]{borthakur2019dynamics}%
  \BibitemOpen
  \bibfield  {author} {\bibinfo {author} {\bibfnamefont {M.~P.}\ \bibnamefont
  {Borthakur}}, \bibinfo {author} {\bibfnamefont {G.}~\bibnamefont {Biswas}},
  \bibinfo {author} {\bibfnamefont {D.}~\bibnamefont {Bandyopadhyay}}, \ and\
  \bibinfo {author} {\bibfnamefont {K.~C.}\ \bibnamefont {Sahu}},\ }\bibfield
  {title} {\enquote {\bibinfo {title} {Dynamics of an arched liquid jet under
  the influence of gravity},}\ }\href@noop {} {\bibfield  {journal} {\bibinfo
  {journal} {Eur. J. Mech. B Fluids}\ }\textbf {\bibinfo {volume} {74}},\
  \bibinfo {pages} {1--9} (\bibinfo {year} {2019})}\BibitemShut {NoStop}%
\bibitem [{\citenamefont {Birouk}\ and\ \citenamefont
  {Lekic}(2009)}]{birouk2009liquid}%
  \BibitemOpen
  \bibfield  {author} {\bibinfo {author} {\bibfnamefont {M.}~\bibnamefont
  {Birouk}}\ and\ \bibinfo {author} {\bibfnamefont {N.}~\bibnamefont {Lekic}},\
  }\bibfield  {title} {\enquote {\bibinfo {title} {Liquid jet breakup in
  quiescent atmosphere: A review},}\ }\href@noop {} {\bibfield  {journal}
  {\bibinfo  {journal} {At. Sprays}\ }\textbf {\bibinfo {volume} {19}},\
  \bibinfo {pages} {501-528} (\bibinfo {year} {2009})}\BibitemShut {NoStop}%
\bibitem [{\citenamefont {Grant}\ and\ \citenamefont
  {Middleman}(1966)}]{grant1966newtonian}%
  \BibitemOpen
  \bibfield  {author} {\bibinfo {author} {\bibfnamefont {R.~P.}\ \bibnamefont
  {Grant}}\ and\ \bibinfo {author} {\bibfnamefont {S.}~\bibnamefont
  {Middleman}},\ }\bibfield  {title} {\enquote {\bibinfo {title} {Newtonian jet
  stability},}\ }\href@noop {} {\bibfield  {journal} {\bibinfo  {journal}
  {AIChE J.}\ }\textbf {\bibinfo {volume} {12}},\ \bibinfo {pages} {669--678}
  (\bibinfo {year} {1966})}\BibitemShut {NoStop}%
\bibitem [{\citenamefont {Srinivasan}\ and\ \citenamefont
  {Sinha}(2024)}]{srinivasan2024primary}%
  \BibitemOpen
  \bibfield  {author} {\bibinfo {author} {\bibfnamefont {B.}~\bibnamefont
  {Srinivasan}}\ and\ \bibinfo {author} {\bibfnamefont {A.}~\bibnamefont
  {Sinha}},\ }\bibfield  {title} {\enquote {\bibinfo {title} {Primary breakup
  of liquid jet—effect of jet velocity profile},}\ }\href@noop {} {\bibfield
  {journal} {\bibinfo  {journal} {Phys. Fluids}\ }\textbf {\bibinfo {volume}
  {36}},\ \bibinfo {pages} {032102} (\bibinfo {year} {2024})}\BibitemShut {NoStop}%
\bibitem [{\citenamefont {Tanase}, \citenamefont {Rasuceanu},\ and\
  \citenamefont {Balan}(2023)}]{tanase2023experimental}%
  \BibitemOpen
  \bibfield  {author} {\bibinfo {author} {\bibfnamefont {N.~O.}\ \bibnamefont
  {Tanase}}, \bibinfo {author} {\bibfnamefont {I.}~\bibnamefont {Rasuceanu}}, \
  and\ \bibinfo {author} {\bibfnamefont {C.}~\bibnamefont {Balan}},\ }\bibfield
   {title} {\enquote {\bibinfo {title} {Experimental and numerical
  investigations of the jetting regime},}\ }in\ \href@noop {} {\emph {\bibinfo
  {booktitle} {IOP Conf. Ser.: Earth Environ. Sci.}}},\ Vol.\ \bibinfo {volume}
  {1136}\ (\bibinfo {organization} {IOP Publishing},\ \bibinfo {year} {2023})\
  p.\ \bibinfo {pages} {012038}\BibitemShut {NoStop}%
\bibitem [{\citenamefont {Kiaoulias}\ \emph {et~al.}(2019)\citenamefont
  {Kiaoulias}, \citenamefont {Travis}, \citenamefont {Moore},\ and\
  \citenamefont {Risha}}]{kiaoulias2019evaluation}%
  \BibitemOpen
  \bibfield  {author} {\bibinfo {author} {\bibfnamefont {D.}~\bibnamefont
  {Kiaoulias}}, \bibinfo {author} {\bibfnamefont {T.}~\bibnamefont {Travis}},
  \bibinfo {author} {\bibfnamefont {J.}~\bibnamefont {Moore}}, \ and\ \bibinfo
  {author} {\bibfnamefont {G.}~\bibnamefont {Risha}},\ }\bibfield  {title}
  {\enquote {\bibinfo {title} {Evaluation of orifice inlet geometries on single
  liquid injectors through cold-flow experiments},}\ }\href@noop {} {\bibfield
  {journal} {\bibinfo  {journal} {Exp. Therm. Fluid Sci.}\ }\textbf {\bibinfo
  {volume} {103}},\ \bibinfo {pages} {78--88} (\bibinfo {year}
  {2019})}\BibitemShut {NoStop}%
\bibitem [{\citenamefont {Sharma}\ and\ \citenamefont
  {Fang}(2014)}]{sharma2014breakup}%
  \BibitemOpen
  \bibfield  {author} {\bibinfo {author} {\bibfnamefont {P.}~\bibnamefont
  {Sharma}}\ and\ \bibinfo {author} {\bibfnamefont {T.}~\bibnamefont {Fang}},\
  }\bibfield  {title} {\enquote {\bibinfo {title} {Breakup of liquid jets from
  non-circular orifices},}\ }\href@noop {} {\bibfield  {journal} {\bibinfo
  {journal} {Exp. Fluids}\ }\textbf {\bibinfo {volume} {55}},\ \bibinfo {pages}
  {1--17} (\bibinfo {year} {2014})}\BibitemShut {NoStop}%
\bibitem [{\citenamefont {Wang}\ and\ \citenamefont
  {Fang}(2015)}]{wang2015liquid}%
  \BibitemOpen
  \bibfield  {author} {\bibinfo {author} {\bibfnamefont {F.}~\bibnamefont
  {Wang}}\ and\ \bibinfo {author} {\bibfnamefont {T.}~\bibnamefont {Fang}},\
  }\bibfield  {title} {\enquote {\bibinfo {title} {Liquid jet breakup for
  non-circular orifices under low pressures},}\ }\href@noop {} {\bibfield
  {journal} {\bibinfo  {journal} {Int. J. Multiphase Flow}\ }\textbf {\bibinfo
  {volume} {72}},\ \bibinfo {pages} {248--262} (\bibinfo {year}
  {2015})}\BibitemShut {NoStop}%
\bibitem [{\citenamefont {Borthakur}, \citenamefont {Biswas},\ and\
  \citenamefont {Bandyopadhyay}(2017)}]{borthakur2017formation}%
  \BibitemOpen
  \bibfield  {author} {\bibinfo {author} {\bibfnamefont {M.~P.}\ \bibnamefont
  {Borthakur}}, \bibinfo {author} {\bibfnamefont {G.}~\bibnamefont {Biswas}}, \
  and\ \bibinfo {author} {\bibfnamefont {D.}~\bibnamefont {Bandyopadhyay}},\
  }\bibfield  {title} {\enquote {\bibinfo {title} {Formation of liquid drops at
  an orifice and dynamics of pinch-off in liquid jets},}\ }\href@noop {}
  {\bibfield  {journal} {\bibinfo  {journal} {Phys. Rev. E.}\ }\textbf
  {\bibinfo {volume} {96}},\ \bibinfo {pages} {013115} (\bibinfo {year}
  {2017})}\BibitemShut {NoStop}%
\bibitem [{\citenamefont {Rajesh}, \citenamefont {Sakthikumar},\ and\
  \citenamefont {Sivakumar}(2016)}]{rajesh2016interfacial}%
  \BibitemOpen
  \bibfield  {author} {\bibinfo {author} {\bibfnamefont {K.~R.}\ \bibnamefont
  {Rajesh}}, \bibinfo {author} {\bibfnamefont {R.}~\bibnamefont {Sakthikumar}},
  \ and\ \bibinfo {author} {\bibfnamefont {D.}~\bibnamefont {Sivakumar}},\
  }\bibfield  {title} {\enquote {\bibinfo {title} {Interfacial oscillation of
  liquid jets discharging from non-circular orifices},}\ }\href@noop {}
  {\bibfield  {journal} {\bibinfo  {journal} {Int. J. Multiphase Flow}\
  }\textbf {\bibinfo {volume} {87}},\ \bibinfo {pages} {1--8} (\bibinfo {year}
  {2016})}\BibitemShut {NoStop}%
\bibitem [{\citenamefont {Geng}\ \emph {et~al.}(2020)\citenamefont {Geng},
  \citenamefont {Wang}, \citenamefont {Wang},\ and\ \citenamefont
  {Li}}]{geng2020effect}%
  \BibitemOpen
  \bibfield  {author} {\bibinfo {author} {\bibfnamefont {L.}~\bibnamefont
  {Geng}}, \bibinfo {author} {\bibfnamefont {Y.}~\bibnamefont {Wang}}, \bibinfo
  {author} {\bibfnamefont {Y.}~\bibnamefont {Wang}}, \ and\ \bibinfo {author}
  {\bibfnamefont {H.}~\bibnamefont {Li}},\ }\bibfield  {title} {\enquote
  {\bibinfo {title} {Effect of the injection pressure and orifice diameter on
  the spray characteristics of biodiesel},}\ }\href@noop {} {\bibfield
  {journal} {\bibinfo  {journal} {J. Traffic Transp. Eng. (Engl. Ed.)}\
  }\textbf {\bibinfo {volume} {7}},\ \bibinfo {pages} {331--339} (\bibinfo
  {year} {2020})}\BibitemShut {NoStop}%
\bibitem [{\citenamefont {Rajesh}\ \emph {et~al.}(2023)\citenamefont {Rajesh},
  \citenamefont {Kulkarni}, \citenamefont {Vankeswaram}, \citenamefont
  {Sakthikumar},\ and\ \citenamefont {Deivandren}}]{rajesh2023drop}%
  \BibitemOpen
  \bibfield  {author} {\bibinfo {author} {\bibfnamefont {K.~R.}\ \bibnamefont
  {Rajesh}}, \bibinfo {author} {\bibfnamefont {V.}~\bibnamefont {Kulkarni}},
  \bibinfo {author} {\bibfnamefont {S.~K.}\ \bibnamefont {Vankeswaram}},
  \bibinfo {author} {\bibfnamefont {R.}~\bibnamefont {Sakthikumar}}, \ and\
  \bibinfo {author} {\bibfnamefont {S.}~\bibnamefont {Deivandren}},\ }\bibfield
   {title} {\enquote {\bibinfo {title} {Drop size characteristics of sprays
  emanating from circular and non-circular orifices in the atomization
  regime},}\ }\href@noop {} {\bibfield  {journal} {\bibinfo  {journal} {J.
  Aerosol Sci.}\ }\textbf {\bibinfo {volume} {174}},\ \bibinfo {pages} {106245}
  (\bibinfo {year} {2023})}\BibitemShut {NoStop}%
\bibitem [{\citenamefont {Tadjfar}\ and\ \citenamefont
  {Jaberi}(2019)}]{tadjfar2019effects}%
  \BibitemOpen
  \bibfield  {author} {\bibinfo {author} {\bibfnamefont {M.}~\bibnamefont
  {Tadjfar}}\ and\ \bibinfo {author} {\bibfnamefont {A.}~\bibnamefont
  {Jaberi}},\ }\bibfield  {title} {\enquote {\bibinfo {title} {Effects of
  aspect ratio on the flow development of rectangular liquid jets issued into
  stagnant air},}\ }\href@noop {} {\bibfield  {journal} {\bibinfo  {journal}
  {Int. J. Multiphase Flow}\ }\textbf {\bibinfo {volume} {115}},\ \bibinfo
  {pages} {144--157} (\bibinfo {year} {2019})}\BibitemShut {NoStop}%
\bibitem [{\citenamefont {Morad}, \citenamefont {Nasiri},\ and\ \citenamefont
  {Amini}(2020)}]{morad2020axis}%
  \BibitemOpen
  \bibfield  {author} {\bibinfo {author} {\bibfnamefont {M.~R.}\ \bibnamefont
  {Morad}}, \bibinfo {author} {\bibfnamefont {M.}~\bibnamefont {Nasiri}}, \
  and\ \bibinfo {author} {\bibfnamefont {G.}~\bibnamefont {Amini}},\ }\bibfield
   {title} {\enquote {\bibinfo {title} {Axis-switching and breakup of
  rectangular liquid jets},}\ }\href@noop {} {\bibfield  {journal} {\bibinfo
  {journal} {Int. J. Multiphase Flow}\ }\textbf {\bibinfo {volume} {126}},\
  \bibinfo {pages} {103242} (\bibinfo {year} {2020})}\BibitemShut {NoStop}%
\bibitem [{\citenamefont {Jaberi}\ and\ \citenamefont
  {Tadjfar}(2019)}]{jaberi2019wavelength}%
  \BibitemOpen
  \bibfield  {author} {\bibinfo {author} {\bibfnamefont {A.}~\bibnamefont
  {Jaberi}}\ and\ \bibinfo {author} {\bibfnamefont {M.}~\bibnamefont
  {Tadjfar}},\ }\bibfield  {title} {\enquote {\bibinfo {title} {Wavelength and
  frequency of axis-switching phenomenon formed over rectangular and elliptical
  liquid jets},}\ }\href@noop {} {\bibfield  {journal} {\bibinfo  {journal}
  {Int. J. Multiphase Flow}\ }\textbf {\bibinfo {volume} {119}},\ \bibinfo
  {pages} {144--154} (\bibinfo {year} {2019})}\BibitemShut {NoStop}%
\bibitem [{\citenamefont {Kooij}\ \emph {et~al.}(2018)\citenamefont {Kooij},
  \citenamefont {Sijs}, \citenamefont {Denn}, \citenamefont {Villermaux},\ and\
  \citenamefont {Bonn}}]{kooij2018determines}%
  \BibitemOpen
  \bibfield  {author} {\bibinfo {author} {\bibfnamefont {S.}~\bibnamefont
  {Kooij}}, \bibinfo {author} {\bibfnamefont {R.}~\bibnamefont {Sijs}},
  \bibinfo {author} {\bibfnamefont {M.~M.}\ \bibnamefont {Denn}}, \bibinfo
  {author} {\bibfnamefont {E.}~\bibnamefont {Villermaux}}, \ and\ \bibinfo
  {author} {\bibfnamefont {D.}~\bibnamefont {Bonn}},\ }\bibfield  {title}
  {\enquote {\bibinfo {title} {What determines the drop size in sprays?}}\
  }\href@noop {} {\bibfield  {journal} {\bibinfo  {journal} {Phys. Rev. X}\
  }\textbf {\bibinfo {volume} {8}},\ \bibinfo {pages} {031019} (\bibinfo {year}
  {2018})}\BibitemShut {NoStop}%
\bibitem [{\citenamefont {Roth}\ \emph {et~al.}(2021)\citenamefont {Roth},
  \citenamefont {Frantz}, \citenamefont {Chaze}, \citenamefont {Corber},\ and\
  \citenamefont {Berrocal}}]{roth2021high}%
  \BibitemOpen
  \bibfield  {author} {\bibinfo {author} {\bibfnamefont {A.}~\bibnamefont
  {Roth}}, \bibinfo {author} {\bibfnamefont {D.}~\bibnamefont {Frantz}},
  \bibinfo {author} {\bibfnamefont {W.}~\bibnamefont {Chaze}}, \bibinfo
  {author} {\bibfnamefont {A.}~\bibnamefont {Corber}}, \ and\ \bibinfo {author}
  {\bibfnamefont {E.}~\bibnamefont {Berrocal}},\ }\bibfield  {title} {\enquote
  {\bibinfo {title} {High-speed imaging database of water jet disintegration
  part i: Quantitative imaging using liquid laser-induced fluorescence},}\
  }\href@noop {} {\bibfield  {journal} {\bibinfo  {journal} {Int. J. Multiphase
  Flow}\ }\textbf {\bibinfo {volume} {145}},\ \bibinfo {pages} {103641}
  (\bibinfo {year} {2021})}\BibitemShut {NoStop}%
\bibitem [{\citenamefont {Rezayat}, \citenamefont {Farshchi},\ and\
  \citenamefont {Berrocal}(2021)}]{rezayat2021high}%
  \BibitemOpen
  \bibfield  {author} {\bibinfo {author} {\bibfnamefont {S.}~\bibnamefont
  {Rezayat}}, \bibinfo {author} {\bibfnamefont {M.}~\bibnamefont {Farshchi}}, \
  and\ \bibinfo {author} {\bibfnamefont {E.}~\bibnamefont {Berrocal}},\
  }\bibfield  {title} {\enquote {\bibinfo {title} {High-speed imaging database
  of water jet disintegration part ii: Temporal analysis of the primary
  breakup},}\ }\href@noop {} {\bibfield  {journal} {\bibinfo  {journal} {Int.
  J. Multiphase Flow}\ }\textbf {\bibinfo {volume} {145}},\ \bibinfo {pages}
  {103807} (\bibinfo {year} {2021})}\BibitemShut {NoStop}%
\bibitem [{\citenamefont {Kang}\ \emph {et~al.}(2023)\citenamefont {Kang},
  \citenamefont {Yu}, \citenamefont {Liu},\ and\ \citenamefont
  {Tao}}]{kang2023effect}%
  \BibitemOpen
  \bibfield  {author} {\bibinfo {author} {\bibfnamefont {T.}~\bibnamefont
  {Kang}}, \bibinfo {author} {\bibfnamefont {Q.}~\bibnamefont {Yu}}, \bibinfo
  {author} {\bibfnamefont {Z.}~\bibnamefont {Liu}}, \ and\ \bibinfo {author}
  {\bibfnamefont {S.}~\bibnamefont {Tao}},\ }\bibfield  {title} {\enquote
  {\bibinfo {title} {Effect of velocity and radius distribution on jet
  breakup},}\ }\href@noop {} {\bibfield  {journal} {\bibinfo  {journal} {Phys.
  Fluids}\ }\textbf {\bibinfo {volume} {35}},\ \bibinfo {pages}
  {014103} (\bibinfo {year}
  {2023})}\BibitemShut {NoStop}%
\bibitem [{\citenamefont {Zhan}\ \emph {et~al.}(2020)\citenamefont {Zhan},
  \citenamefont {Kuwata}, \citenamefont {Maruyama}, \citenamefont {Okawa},
  \citenamefont {Enoki}, \citenamefont {Aoyagi},\ and\ \citenamefont
  {Takata}}]{zhan2020effects}%
  \BibitemOpen
  \bibfield  {author} {\bibinfo {author} {\bibfnamefont {Y.}~\bibnamefont
  {Zhan}}, \bibinfo {author} {\bibfnamefont {Y.}~\bibnamefont {Kuwata}},
  \bibinfo {author} {\bibfnamefont {K.}~\bibnamefont {Maruyama}}, \bibinfo
  {author} {\bibfnamefont {T.}~\bibnamefont {Okawa}}, \bibinfo {author}
  {\bibfnamefont {K.}~\bibnamefont {Enoki}}, \bibinfo {author} {\bibfnamefont
  {M.}~\bibnamefont {Aoyagi}}, \ and\ \bibinfo {author} {\bibfnamefont
  {T.}~\bibnamefont {Takata}},\ }\bibfield  {title} {\enquote {\bibinfo {title}
  {Effects of surface tension and viscosity on liquid jet breakup},}\
  }\href@noop {} {\bibfield  {journal} {\bibinfo  {journal} {Exp. Therm. Fluid
  Sci.}\ }\textbf {\bibinfo {volume} {112}},\ \bibinfo {pages} {109953}
  (\bibinfo {year} {2020})}\BibitemShut {NoStop}%
\bibitem [{\citenamefont {Farvardin}\ and\ \citenamefont
  {Dolatabadi}(2013)}]{farvardin2013numerical}%
  \BibitemOpen
  \bibfield  {author} {\bibinfo {author} {\bibfnamefont {E.}~\bibnamefont
  {Farvardin}}\ and\ \bibinfo {author} {\bibfnamefont {A.}~\bibnamefont
  {Dolatabadi}},\ }\bibfield  {title} {\enquote {\bibinfo {title} {Numerical
  simulation of the breakup of elliptical liquid jet in still air},}\
  }\href@noop {} {\bibfield  {journal} {\bibinfo  {journal} {J. Fluids Eng.
  Trans. ASME}\ }\textbf {\bibinfo {volume} {135}},\ \bibinfo {pages} {071302}
  (\bibinfo {year} {2013})}\BibitemShut {NoStop}%
\bibitem [{\citenamefont {Li}\ \emph {et~al.}(2019)\citenamefont {Li},
  \citenamefont {Li}, \citenamefont {Xiao}, \citenamefont {Li},\ and\
  \citenamefont {Zhu}}]{li2019experimental}%
  \BibitemOpen
  \bibfield  {author} {\bibinfo {author} {\bibfnamefont {C.}~\bibnamefont
  {Li}}, \bibinfo {author} {\bibfnamefont {C.}~\bibnamefont {Li}}, \bibinfo
  {author} {\bibfnamefont {F.}~\bibnamefont {Xiao}}, \bibinfo {author}
  {\bibfnamefont {Q.}~\bibnamefont {Li}}, \ and\ \bibinfo {author}
  {\bibfnamefont {Y.}~\bibnamefont {Zhu}},\ }\bibfield  {title} {\enquote
  {\bibinfo {title} {Experimental study of spray characteristics of liquid jets
  in supersonic crossflow},}\ }\href@noop {} {\bibfield  {journal} {\bibinfo
  {journal} {Aerosp. Sci. Technol.}\ }\textbf {\bibinfo {volume} {95}},\
  \bibinfo {pages} {105426} (\bibinfo {year} {2019})}\BibitemShut {NoStop}%
\bibitem [{\citenamefont {Chakraborty}, \citenamefont {Sahu},\ and\
  \citenamefont {Maurya}(2022)}]{chakraborty2022effect}%
  \BibitemOpen
  \bibfield  {author} {\bibinfo {author} {\bibfnamefont {A.}~\bibnamefont
  {Chakraborty}}, \bibinfo {author} {\bibfnamefont {S.}~\bibnamefont {Sahu}}, \
  and\ \bibinfo {author} {\bibfnamefont {D.}~\bibnamefont {Maurya}},\
  }\bibfield  {title} {\enquote {\bibinfo {title} {Effect of orifice size on
  liquid breakup dynamics and spray characteristics in slinger atomizers},}\
  }\href@noop {} {\bibfield  {journal} {\bibinfo  {journal} {Proc. Inst. Mech.
  Eng. A: J. Power Energy}\ }\textbf {\bibinfo {volume} {236}},\ \bibinfo
  {pages} {1158--1170} (\bibinfo {year} {2022})}\BibitemShut {NoStop}%
\bibitem [{\citenamefont {Reitz}(1978)}]{reitz1978atomization}%
  \BibitemOpen
  \bibfield  {author} {\bibinfo {author} {\bibfnamefont {R.~D.}\ \bibnamefont
  {Reitz}},\ }\href@noop {} {\emph {\bibinfo {title} {Atomization and other
  breakup regimes of a liquid jet.}}}\ (\bibinfo  {publisher} {Princeton
  University},\ \bibinfo {year} {1978})\BibitemShut {NoStop}%
\bibitem [{\citenamefont {Richards}, \citenamefont {Lenhoff},\ and\
  \citenamefont {Beris}(1994)}]{richards1994dynamic}%
  \BibitemOpen
  \bibfield  {author} {\bibinfo {author} {\bibfnamefont {J.~R.}\ \bibnamefont
  {Richards}}, \bibinfo {author} {\bibfnamefont {A.~M.}\ \bibnamefont
  {Lenhoff}}, \ and\ \bibinfo {author} {\bibfnamefont {A.~N.}\ \bibnamefont
  {Beris}},\ }\bibfield  {title} {\enquote {\bibinfo {title} {Dynamic breakup
  of liquid--liquid jets},}\ }\href@noop {} {\bibfield  {journal} {\bibinfo
  {journal} {Phys. Fluids}\ }\textbf {\bibinfo {volume} {6}},\ \bibinfo {pages}
  {2640--2655} (\bibinfo {year} {1994})}\BibitemShut {NoStop}%
\bibitem [{\citenamefont {Machicoane}, \citenamefont {Osuna-Orozco},\ and\
  \citenamefont {Aliseda}(2023)}]{machicoane2023regimes}%
  \BibitemOpen
  \bibfield  {author} {\bibinfo {author} {\bibfnamefont {N.}~\bibnamefont
  {Machicoane}}, \bibinfo {author} {\bibfnamefont {R.}~\bibnamefont
  {Osuna-Orozco}}, \ and\ \bibinfo {author} {\bibfnamefont {A.}~\bibnamefont
  {Aliseda}},\ }\bibfield  {title} {\enquote {\bibinfo {title} {Regimes of the
  length of a laminar liquid jet fragmented by a gas co-flow},}\ }\href@noop {}
  {\bibfield  {journal} {\bibinfo  {journal} {Int. J. Multiphase Flow}\
  }\textbf {\bibinfo {volume} {165}},\ \bibinfo {pages} {104475} (\bibinfo
  {year} {2023})}\BibitemShut {NoStop}%
\bibitem [{\citenamefont {Dunand}, \citenamefont {Carreau},\ and\ \citenamefont
  {Roger}(2005)}]{dunand2005liquid}%
  \BibitemOpen
  \bibfield  {author} {\bibinfo {author} {\bibfnamefont {A.}~\bibnamefont
  {Dunand}}, \bibinfo {author} {\bibfnamefont {J.-L.}\ \bibnamefont {Carreau}},
  \ and\ \bibinfo {author} {\bibfnamefont {F.}~\bibnamefont {Roger}},\
  }\bibfield  {title} {\enquote {\bibinfo {title} {Liquid jet breakup and
  atomization by annular swirling gas jet},}\ }\href@noop {} {\bibfield
  {journal} {\bibinfo  {journal} {At. Sprays}\ }\textbf {\bibinfo {volume}
  {15}},\ \bibinfo {pages} {223-247} (\bibinfo {year} {2005})}\BibitemShut {NoStop}%
\bibitem [{\citenamefont {Daskiran}\ \emph {et~al.}(2022)\citenamefont
  {Daskiran}, \citenamefont {Xue}, \citenamefont {Cui}, \citenamefont {Katz},\
  and\ \citenamefont {Boufadel}}]{daskiran2022impact}%
  \BibitemOpen
  \bibfield  {author} {\bibinfo {author} {\bibfnamefont {C.}~\bibnamefont
  {Daskiran}}, \bibinfo {author} {\bibfnamefont {X.}~\bibnamefont {Xue}},
  \bibinfo {author} {\bibfnamefont {F.}~\bibnamefont {Cui}}, \bibinfo {author}
  {\bibfnamefont {J.}~\bibnamefont {Katz}}, \ and\ \bibinfo {author}
  {\bibfnamefont {M.~C.}\ \bibnamefont {Boufadel}},\ }\bibfield  {title}
  {\enquote {\bibinfo {title} {Impact of a jet orifice on the hydrodynamics and
  the oil droplet size distribution},}\ }\href@noop {} {\bibfield  {journal}
  {\bibinfo  {journal} {Int. J. Multiphase Flow}\ }\textbf {\bibinfo {volume}
  {147}},\ \bibinfo {pages} {103921} (\bibinfo {year} {2022})}\BibitemShut
  {NoStop}%
\bibitem [{\citenamefont {Kovalchuk}\ and\ \citenamefont
  {Simmons}(2018)}]{kovalchuk2018effect}%
  \BibitemOpen
  \bibfield  {author} {\bibinfo {author} {\bibfnamefont {N.~M.}\ \bibnamefont
  {Kovalchuk}}\ and\ \bibinfo {author} {\bibfnamefont {M.~J.}\ \bibnamefont
  {Simmons}},\ }\bibfield  {title} {\enquote {\bibinfo {title} {Effect of
  soluble surfactant on regime transitions at drop formation},}\ }\href@noop {}
  {\bibfield  {journal} {\bibinfo  {journal} {Colloids Surf. A: Physicochem.
  Eng. Asp.}\ }\textbf {\bibinfo {volume} {545}},\ \bibinfo {pages} {1--7}
  (\bibinfo {year} {2018})}\BibitemShut {NoStop}%
\bibitem [{\citenamefont {Furbank}\ and\ \citenamefont
  {Morris}(2004)}]{furbank2004experimental}%
  \BibitemOpen
  \bibfield  {author} {\bibinfo {author} {\bibfnamefont {R.~J.}\ \bibnamefont
  {Furbank}}\ and\ \bibinfo {author} {\bibfnamefont {J.~F.}\ \bibnamefont
  {Morris}},\ }\bibfield  {title} {\enquote {\bibinfo {title} {An experimental
  study of particle effects on drop formation},}\ }\href@noop {} {\bibfield
  {journal} {\bibinfo  {journal} {Phys. Fluids}\ }\textbf {\bibinfo {volume}
  {16}},\ \bibinfo {pages} {1777--1790} (\bibinfo {year} {2004})}\BibitemShut
  {NoStop}%
\bibitem [{\citenamefont {Rayleigh}(1879)}]{rayleigh1879capillary}%
  \BibitemOpen
  \bibfield  {author} {\bibinfo {author} {\bibfnamefont {L.}~\bibnamefont
  {Rayleigh}},\ }\bibfield  {title} {\enquote {\bibinfo {title} {On the
  capillary phenomena of jets},}\ }\href@noop {} {\bibfield  {journal}
  {\bibinfo  {journal} {Proc. R. Soc. Lond.}\ }\textbf {\bibinfo {volume}
  {29}},\ \bibinfo {pages} {71--97} (\bibinfo {year} {1879})}\BibitemShut
  {NoStop}%
\bibitem [{\citenamefont {Kalaaji}\ \emph {et~al.}(2003)\citenamefont
  {Kalaaji}, \citenamefont {Lopez}, \citenamefont {Attan{\'e}},\ and\
  \citenamefont {Soucemarianadin}}]{kalaaji2003breakup}%
  \BibitemOpen
  \bibfield  {author} {\bibinfo {author} {\bibfnamefont {A.}~\bibnamefont
  {Kalaaji}}, \bibinfo {author} {\bibfnamefont {B.}~\bibnamefont {Lopez}},
  \bibinfo {author} {\bibfnamefont {P.}~\bibnamefont {Attan{\'e}}}, \ and\
  \bibinfo {author} {\bibfnamefont {A.}~\bibnamefont {Soucemarianadin}},\
  }\bibfield  {title} {\enquote {\bibinfo {title} {Breakup length of forced
  liquid jets},}\ }\href@noop {} {\bibfield  {journal} {\bibinfo  {journal}
  {Phys. Fluids}\ }\textbf {\bibinfo {volume} {15}},\ \bibinfo {pages}
  {2469--2479} (\bibinfo {year} {2003})}\BibitemShut {NoStop}%
\bibitem [{\citenamefont {Meister}\ and\ \citenamefont
  {Scheele}(1969)}]{meister1969prediction}%
  \BibitemOpen
  \bibfield  {author} {\bibinfo {author} {\bibfnamefont {B.~J.}\ \bibnamefont
  {Meister}}\ and\ \bibinfo {author} {\bibfnamefont {G.~F.}\ \bibnamefont
  {Scheele}},\ }\bibfield  {title} {\enquote {\bibinfo {title} {Prediction of
  jet length in immiscible liquid systems},}\ }\href@noop {} {\bibfield
  {journal} {\bibinfo  {journal} {AIChE J.}\ }\textbf {\bibinfo {volume}
  {15}},\ \bibinfo {pages} {689--699} (\bibinfo {year} {1969})}\BibitemShut
  {NoStop}%
\bibitem [{\citenamefont {Sun}\ \emph {et~al.}(2024)\citenamefont {Sun},
  \citenamefont {Zhan}, \citenamefont {Okawa}, \citenamefont {Aoyagi},
  \citenamefont {Uchibori},\ and\ \citenamefont {Okano}}]{sun2024effects}%
  \BibitemOpen
  \bibfield  {author} {\bibinfo {author} {\bibfnamefont {G.}~\bibnamefont
  {Sun}}, \bibinfo {author} {\bibfnamefont {Y.}~\bibnamefont {Zhan}}, \bibinfo
  {author} {\bibfnamefont {T.}~\bibnamefont {Okawa}}, \bibinfo {author}
  {\bibfnamefont {M.}~\bibnamefont {Aoyagi}}, \bibinfo {author} {\bibfnamefont
  {A.}~\bibnamefont {Uchibori}}, \ and\ \bibinfo {author} {\bibfnamefont
  {Y.}~\bibnamefont {Okano}},\ }\bibfield  {title} {\enquote {\bibinfo {title}
  {Effects of nozzle orifice shape on jet breakup and splashing during liquid
  jet impact onto a horizontal plate},}\ }\href@noop {} {\bibfield  {journal}
  {\bibinfo  {journal} {Exp. Therm. Fluid Sci.}\ }\textbf {\bibinfo {volume}
  {151}},\ \bibinfo {pages} {111095} (\bibinfo {year} {2024})}\BibitemShut
  {NoStop}%
\bibitem [{\citenamefont {Mansour}\ and\ \citenamefont
  {Chigier}(1991)}]{mansour1991dynamic}%
  \BibitemOpen
  \bibfield  {author} {\bibinfo {author} {\bibfnamefont {A.}~\bibnamefont
  {Mansour}}\ and\ \bibinfo {author} {\bibfnamefont {N.}~\bibnamefont
  {Chigier}},\ }\bibfield  {title} {\enquote {\bibinfo {title} {Dynamic
  behavior of liquid sheets},}\ }\href@noop {} {\bibfield  {journal} {\bibinfo
  {journal} {Phys Fluids A, Fluid Dyn}\ }\textbf {\bibinfo {volume} {3}},\
  \bibinfo {pages} {2971--2980} (\bibinfo {year} {1991})}\BibitemShut {NoStop}%
\bibitem [{\citenamefont {Javadi}\ \emph {et~al.}(2013)\citenamefont {Javadi},
  \citenamefont {Eggers}, \citenamefont {Bonn}, \citenamefont {Habibi},\ and\
  \citenamefont {Ribe}}]{javadi2013delayed}%
  \BibitemOpen
  \bibfield  {author} {\bibinfo {author} {\bibfnamefont {A.}~\bibnamefont
  {Javadi}}, \bibinfo {author} {\bibfnamefont {J.}~\bibnamefont {Eggers}},
  \bibinfo {author} {\bibfnamefont {D.}~\bibnamefont {Bonn}}, \bibinfo {author}
  {\bibfnamefont {M.}~\bibnamefont {Habibi}}, \ and\ \bibinfo {author}
  {\bibfnamefont {N.~M.}\ \bibnamefont {Ribe}},\ }\bibfield  {title} {\enquote
  {\bibinfo {title} {Delayed capillary breakup of falling viscous jets},}\
  }\href@noop {} {\bibfield  {journal} {\bibinfo  {journal} {Phys. Rev. Lett.}\
  }\textbf {\bibinfo {volume} {110}},\ \bibinfo {pages} {144501} (\bibinfo
  {year} {2013})}\BibitemShut {NoStop}%
\bibitem [{\citenamefont {Kirar}\ \emph
  {et~al.}(2022{\natexlab{b}})\citenamefont {Kirar}, \citenamefont {Pokale},
  \citenamefont {Sahu}, \citenamefont {Ray},\ and\ \citenamefont
  {Biswas}}]{kirar2022influence}%
  \BibitemOpen
  \bibfield  {author} {\bibinfo {author} {\bibfnamefont {P.~K.}\ \bibnamefont
  {Kirar}}, \bibinfo {author} {\bibfnamefont {S.~D.}\ \bibnamefont {Pokale}},
  \bibinfo {author} {\bibfnamefont {K.~C.}\ \bibnamefont {Sahu}}, \bibinfo
  {author} {\bibfnamefont {B.}~\bibnamefont {Ray}}, \ and\ \bibinfo {author}
  {\bibfnamefont {G.}~\bibnamefont {Biswas}},\ }\bibfield  {title} {\enquote
  {\bibinfo {title} {Influence of the interaction of capillary waves on the
  dynamics of two drops falling side-by-side on a liquid pool},}\ }\href@noop
  {} {\bibfield  {journal} {\bibinfo  {journal} {Phys. Fluids}\ }\textbf
  {\bibinfo {volume} {34}}, \ \bibinfo {pages} {112114} (\bibinfo {year} {2022}{\natexlab{b}})}\BibitemShut
  {NoStop}%
\bibitem [{\citenamefont {Balla}, \citenamefont {Tripathi},\ and\ \citenamefont
  {Sahu}(2019)}]{balla2019shape}%
  \BibitemOpen
  \bibfield  {author} {\bibinfo {author} {\bibfnamefont {M.}~\bibnamefont
  {Balla}}, \bibinfo {author} {\bibfnamefont {M.~K.}\ \bibnamefont {Tripathi}},
  \ and\ \bibinfo {author} {\bibfnamefont {K.~C.}\ \bibnamefont {Sahu}},\
  }\bibfield  {title} {\enquote {\bibinfo {title} {Shape oscillations of a
  nonspherical water droplet},}\ }\href@noop {} {\bibfield  {journal} {\bibinfo
   {journal} {Phys. Rev. E.}\ }\textbf {\bibinfo {volume} {99}},\ \bibinfo
  {pages} {023107} (\bibinfo {year} {2019})}\BibitemShut {NoStop}%
\bibitem [{\citenamefont {Pal}\ \emph {et~al.}(2024)\citenamefont {Pal},
  \citenamefont {Sahu}, \citenamefont {De},\ and\ \citenamefont
  {Biswas}}]{pal2024collision}%
  \BibitemOpen
  \bibfield  {author} {\bibinfo {author} {\bibfnamefont {A.~K.}\ \bibnamefont
  {Pal}}, \bibinfo {author} {\bibfnamefont {K.~C.}\ \bibnamefont {Sahu}},
  \bibinfo {author} {\bibfnamefont {S.}~\bibnamefont {De}}, \ and\ \bibinfo
  {author} {\bibfnamefont {G.}~\bibnamefont {Biswas}},\ }\bibfield  {title}
  {\enquote {\bibinfo {title} {Collision of two drops moving in the same
  direction},}\ }\href@noop {} {\bibfield  {journal} {\bibinfo  {journal}
  {Phys. Fluids}\ }\textbf {\bibinfo {volume} {36}}, \ \bibinfo {pages} {012122} (\bibinfo {year}
  {2024})}\BibitemShut {NoStop}%
\bibitem [{\citenamefont {Viswanathan}(2019)}]{viswanathan2019breakup}%
  \BibitemOpen
  \bibfield  {author} {\bibinfo {author} {\bibfnamefont {H.}~\bibnamefont
  {Viswanathan}},\ }\bibfield  {title} {\enquote {\bibinfo {title} {Breakup and
  coalescence of drops during transition from dripping to jetting in a
  {N}ewtonian fluid},}\ }\href@noop {} {\bibfield  {journal} {\bibinfo
  {journal} {Int. J. Multiphase Flow}\ }\textbf {\bibinfo {volume} {112}},\
  \bibinfo {pages} {269--285} (\bibinfo {year} {2019})}\BibitemShut {NoStop}%
\end{thebibliography}

%

\end{document}